\documentclass[12pt]{article}

\usepackage[latin1]{inputenc}
\usepackage{graphicx,graphics,epsfig,color}
\usepackage{wrapfig,rotating}
\usepackage{amssymb,amsmath,array}
\usepackage{cite}

\def\CASCADE{{\sc Cascade}}
\def\PYTHIA{{\sc Pythia}}
\def\pythia{{\sc Pythia}}
\def\herwig{{\sc Herwig}}
\def\powheg{{\sc Powheg}}

\def\lesssim{\ \hbox{\raise 2pt \hbox{$<$} \kern -13pt
                     \lower 3pt \hbox{$\sim$}}\ }
\def\greatersim{\ \hbox{\raise 2pt \hbox{$>$} \kern -13pt
                     \lower 3pt \hbox{$\sim$}}\ }

\newcommand{\alphasb}{\bar{\alpha}_s}

\usepackage{amsmath,amssymb,bm,cite}

\begin{document}


\hspace*{12.9 cm} {\small DESY-10-179} 


\hspace*{12.9 cm} {\small OUTP-10-07-P}

\vspace*{1.0 cm} 

\begin{center}
{\Large   \bf 
Forward-Central  Jet Correlations   at the \\
 Large Hadron Collider 
  }
\end{center} 

\begin{center}
{\large  M.~Deak$^{a}$, F.~Hautmann$^{b}$, H.~Jung$^{c,d}$ and K.~Kutak$^{d}$}\\
\vskip 0.2 cm
{$^{a}$Instituto de F{\' i}sica Te{\' o}rica UAM/CSIC, Facultad  de Ciencias, C-XI,\\  
Universidad Aut{\' o}noma de Madrid, Cantoblanco, Madrid 28049}\\ 
 \vskip 0.2 cm  
{$^{b}$Theoretical Physics Department, 
University of Oxford,    Oxford OX1 3NP }\\    
\vskip 0.2 cm
{$^{c}$Deutsches Elektronen Synchrotron,  D-22603  Hamburg}\\    
\vskip 0.2 cm
{$^{d}$ Elementaire Deeltjes Fysica, Universiteit Antwerpen, B 2020 Antwerpen}   
\end{center}

\vskip 1.0 cm 

\begin{center}
\large {Abstract}
\end{center}

For    high-p$_{\rm{T}}$    forward  processes  at   the Large Hadron Collider  (LHC), 
  QCD  logarithmic  corrections in
the hard transverse momentum  and   in the large rapidity  interval
may both be quantitatively  significant.  The theoretical framework
to resum consistently both kinds of  logarithmic corrections
to higher orders in  perturbation theory  is based on
QCD  high-energy factorization.  We present    numerical 
Monte Carlo applications of  
this method  to  final-state observables  associated with production of 
one forward and one central jet.  By  computing 
  jet correlations in rapidity and azimuth,  we  
    analyze   the role of  corrections to 
the parton-showering chain   from  
large-angle gluon radiation,    and  discuss this in 
relationship with Monte Carlo results modeling 
 interactions 
due to multiple parton chains.

\vskip 1.0 cm

\section{Introduction}
\label{sec:intro}

Physics in the forward region at hadron colliders 
is traditionally dominated by soft particle production. With the advent of the LHC, 
forward physics  phenomenology turns  into  a  largely new 
field~\cite{grothe,ajaltouni,denterria}   involving 
both soft and hard production processes, 
because of  the phase space opening up at  high center-of-mass energies.  
Owing to the unprecedented reach in rapidity 
 of  the experimental instrumentation,  
it becomes possible to carry out a program of     high-p$_{\rm{T}}$ 
 physics in the forward region.  
 
Forward jet  production   enters 
the LHC physics program   in  an  essential 
way    both for QCD studies and for new particle searches, e.g. 
in  vector boson fusion search channels for  the 
Higgs boson~\cite{vbf-cms,vbf-atlas}. 
Another area  of  potential interest  in forward physics employs  
  near-beam proton taggers~\cite{albrow_review}:  this  will enable  
studies to be made in the 
central     high-p$_{\rm{T}}$ 
production mode with forward protons, which can  be used 
 for  both standard-candle~\cite{cep_james_etal} and 
discovery physics~\cite{fwdmssm}. 
In addition to collider physics applications, measurements of 
forward particle production 
at the LHC  
will serve as  input   to   the modeling of high-energy air showers in cosmic 
ray experiments~\cite{engel}.

The forward  production  of     high-p$_{\rm{T}}$     particles brings jet physics into a 
  region  characterized  by  multiple energy scales and   
asymmetric parton kinematics. 
  In this multi-scale region it  is compelling 
to ask~\cite{ajaltouni,michel_font} 
whether fixed-order next-to-leading  calculations reliably describe the 
production process  or   significant contributions arise  beyond 
fixed  order  which call for 
perturbative QCD resummations.  
  The   early  observation~\cite{muenav}  of  
potentially large   logarithmic corrections to jet production
   at asymptotically high  energies 
  has given rise to an ample literature  
of  calculations based on 
the BFKL equation~\cite{webetal99,orrsti,mn_pheno,wallon10}. 
On the other hand,  at collider energies  
both logarithmic corrections   in  the large  rapidity  interval  
(of  high-energy type)  
and logarithmic corrections  in  
the hard transverse momentum (of collinear type)
are phenomenologically important.   The theoretical framework 
to resum   consistently 
both  kinds of logarithmic corrections  in QCD  
 perturbation theory    is based on   high-energy    factorization   
 at  fixed transverse momentum~\cite{hef}.  
This  factorization program is carried through  in~\cite{jhep09}  
for forward    jet hadroproduction. 

In this  paper we present the application   of the results~\cite{jhep09}  to 
the study of jet 
correlations  for    production of 
 one forward and one central jet at the LHC.  
 The case of forward-backward jets will be  examined in a forthcoming article. 
 We propose that measurements 
 of    hadronic final-state  observables associated with forward-central 
correlations can provide, 
  starting with the data already  taken at the LHC, a sensitive probe of how well 
  QCD multiple radiation is taken into account in the Monte Carlo event 
  generators  to be used  for   analyses of experimental data in the   
  forward  region.  The results of such investigations  can serve to 
estimate the size of  backgrounds from  QCD radiation 
between  jets   at  large  rapidity separations 
  for  Higgs  boson searches  in 
vector boson fusion channels.  

Besides the contribution of  the higher-order  radiative  corrections taken 
into account  via the results of~\cite{jhep09},  the need for     
realistic  Monte Carlo simulations     of 
 forward particle production    raises the question of whether  
non-negligible effects may come from 
  multiple parton interactions.  Such multiple 
 interactions are modeled in  parton-shower   event generators 
used for  Monte Carlo  simulation of  final states  at the 
LHC~\cite{pz_perugia,rdf1,Sjostrand:2006za,giese}, and form the 
subject of a number of   current 
efforts~\cite{blok,strik-vogel,rog-strik,calu-trel,wiede_mpi,maina_mpi,berger_mpi,gaunt} 
to construct  approaches capable of describing 
  multiple parton scatterings.  
 In this paper 
 we investigate multi-parton interaction effects  for    forward-central 
 jet correlations within the 
 model~\cite{pz_perugia,Sjostrand:2006za}. 
  We  observe that, compared  
to   the production of multiple jets  in 
the single-scattering  picture~\cite{jhep09}, 
the multi-parton mechanism~\cite{pz_perugia,Sjostrand:2006za}    
shifts   a significant  amount of gluon emissions  
to larger values of the longitudinal 
momentum fraction x  in the initial-state decay chains, because less energy is available 
to each of the  sequential  parton chains.   
This results into     differences    
  in the  shapes  of  the  forward-central  jet 
 correlations  in the  azimuth and rapidity  plane between the single-chain and 
 multiple-chain mechanisms for multi-jet production.   
This  can  be investigated at the    LHC    also  via  
measurements  of particle and energy flow  associated with forward production: we leave the 
study of this to  future work.

The paper is organized as follows. 
In Sec.~\ref{secfrom}  we  give a   concise   discussion of  the  high-energy QCD 
 dynamics  underlying  the hadroproduction of forward jets, based 
on the  results of~\cite{jhep09}.    We describe the  high-energy 
 factorized form  of the   forward jet 
cross section that  is to be coupled  to parton showering in order to achieve 
a full description of the associated  hadronic  final states. 
In Sec.~\ref{sec:u-quark}  we discuss  aspects of the  initial-state 
parton showers relevant to  forward hadroproduction.  In particular we 
consider  a method to implement  parton 
 branching  at  transverse-momentum 
dependent   level  not only for 
gluon-initiated channels in  the  backwards evolution 
 but also for  quark-initiated channels. 
In Sec.~\ref{sec:f-jet-lhc}  
we  present  results of  numerical Monte Carlo  calculations   for 
transverse momentum and pseudorapidity spectra and for  
 forward-central   correlations in azimuthal angle and pseudorapidity.  
We  compare  single versus multiple parton interactions, and propose various 
measurements of   forward jet   observables  at the LHC. 
  We give concluding remarks   in  Sec.~\ref{sec:conc}.

\section{Hadroproduction of forward jets} 
\label{secfrom}

This section summarizes results from~\cite{jhep09}. In particular we 
discuss  the physical picture underlying the 
factorization formula that will  be   
used  for  numerical calculations in later sections.  

\begin{figure}[htb]
\vspace{45mm}
\includegraphics{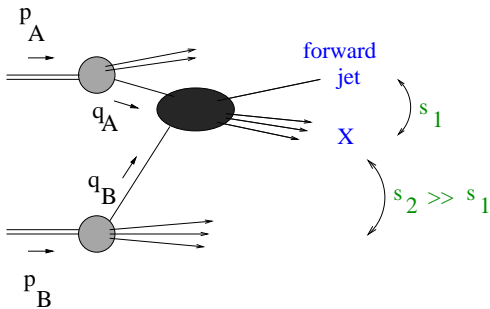}
\caption{\it Jet  production in the forward rapidity region 
in  hadron-hadron collisions.} 
\label{fig:forwpicture}
\end{figure}

  Consider 
 the hadroproduction of a forward jet associated with a hard final  
 state $X$, as depicted in  Fig.~\ref{fig:forwpicture}. 
 The kinematics 
  of the process     is  characterized 
by the  large  ratio  of sub-energies  $s_2  /  s_1 \gg 1 $   
 and  highly asymmetric longitudinal momenta in the partonic initial 
  state   ($x_A \to 1$, $x_B \to 0$).

\begin{figure}[htb]
\vspace{40mm}
\includegraphics{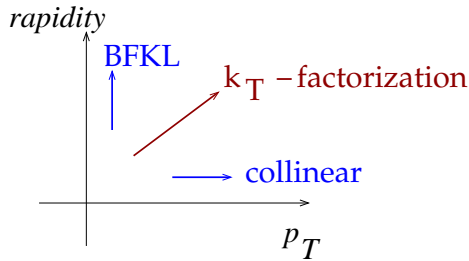}
\caption{\it QCD  radiative contributions  to forward jet  production in  
the rapidity and  transverse momentum plane.} 
\label{fig:ypt}  
\end{figure}

The   presence of multiple   large-momentum scales  
in the  LHC forward   kinematics~\cite{ajaltouni,proceed09}  implies that   
  realistic  phenomenology  of      hadronic  jet final states 
  requires taking  into  account    at   higher  order     
 both logarithmic corrections   in  the large  rapidity  interval  
(of  BFKL  type)  
and logarithmic corrections  in  
the hard transverse momentum (of collinear type).  
This can be  achieved   via QCD  factorization at 
fixed transverse momentum~\cite{hef}. 
A pictorial 
representation of QCD radiative contributions  in the 
rapidity and  transverse momentum plane   
is sketched   in  Fig.~\ref{fig:ypt}.  
Note     that k$_{\rm{T}}$-factorization    is valid to single-logarithmic accuracy. 
In particular, it enables   one to  obtain   logarithmically 
enhanced  terms in rapidity  that are   not associated to 
any collinear logarithm;   conversely, 
  collinear singularities   can be taken into account 
  to any logarithmic accuracy~\cite{ch94}.  
This in contrast with  calculations  in  double-logarithmic approximations. 

The different  expansions   in  Fig.~\ref{fig:ypt} 
   correspond to  different  possible ways of  reorganizing the 
QCD perturbation series.     The results of    factorization at fixed 
k$_{\rm{T}}$ 
can be reobtained by  going to sufficiently sub-leading  orders  in either the 
BFKL  expansion or  the  collinear  expansion~\cite{ch94}.  
This applies for instance  to  the 
  transverse-momentum recoil  effects in the collinear case~\cite{cj05}, and 
kinematic effects of  energy conservation  in the BFKL case~\cite{lundx}.

\subsection{Factorization of the jet cross section }
\label{sec:fact}

  Recall that in the case of forward jet leptoproduction~\cite{mueproc90c} 
QCD factorization at fixed transverse momentum allows one to compute the 
high-energy asymptotic coefficients for  the coupling of forward jets to 
deeply inelastic scattering~\cite{hef,forwjetcoeff}.  
Since the early phenomenological studies~\cite{forwdis92}  forward jet 
leptoproduction has been  investigated at HERA, and will play a major role 
at  the proposed  future   lepton facilities~\cite{laycock} (LHeC, EIC). 
We  come back   to  the possible role of  lepton analyses in Sec.~\ref{sec:conc}.

\begin{figure}[htb]
\vspace{45mm}
\includegraphics{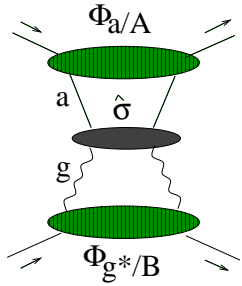}
\caption{\it Factorized structure of the cross section. } 
\label{fig:sec2}
\end{figure}

In the case of  hadroproduction the k$_{\rm{T}}$-factorized form 
 of the  forward  jet  cross section is given in~\cite{jhep09}. This is  
     represented  schematically   in Fig.~\ref{fig:sec2}. 
Initial-state parton configurations  contributing to  
forward  jet  production are asymmetric, 
with the parton in the top subgraph being  probed near  the mass shell and  
large   x,  
while  the parton in  the bottom subgraph is off-shell and small-x. 
The    jet  cross  section differential 
in the final-state   
transverse  momentum 
 $Q_t$  and  azimuthal angle $\varphi$ 
is given  schematically  by  
\begin{equation}
\label{forwsigma}
   {{d   \sigma  } \over 
{ d Q_t^2 d \varphi}} =  \sum_a  \int  \    \phi_{a/A}  \  \otimes \  
 {{d   {\widehat  \sigma}   } \over 
{ d Q_t^2 d \varphi  }}    \  \otimes \   
\phi_{g^*/B}    \;\; , 
\end{equation}
where  
$\otimes$ specifies  a convolution in both longitudinal and transverse momenta, 
$ {\widehat  \sigma} $  is the  hard scattering  cross section,  calculable 
 from  a  suitable off-shell continuation of 
perturbative   matrix elements~\cite{jhep09},  
$ \phi_{a/A} $ is the distribution of parton 
$a$ in hadron $A$ 
obtained from   near-collinear shower evolution, and $ \phi_{g^*/B} $ is  
  the gluon unintegrated distribution in hadron $B$ 
  obtained from non-collinear, 
  transverse momentum  dependent shower evolution. 

\begin{figure}[htb]
\vspace{45mm}
\includegraphics{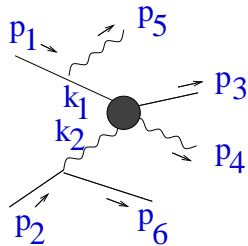}
\caption{\it A typical graph contributing to the  off-shell 
matrix element in the $q g$ channel. } 
\label{fig:sec2graph}
\end{figure}

Fig.~\ref{fig:sec2graph}  shows a typical graph contributing to the 
 off-shell 
matrix element in the $q g$ channel.
By parameterizing the 
exchanged momenta $k_{ 1}$  and $k_{2 }$
in terms of     
purely transverse four-vectors $k_{{\rm{T}} 1}$  and $k_{{\rm{T}} }$ and 
longitudinal 
momentum  fractions $\xi_i$ and $ {\overline \xi}_i$    as  
\begin{equation}
\label{kinek1k2}
p_1 -  p_5 =   k_1 =   \xi_1  p_1 + k_{ {\rm{T}} 1} + {\overline \xi}_1 p_2  \;\;, \;\;\; 
p_2 -  p_6 =   k_2 =   \xi_2  p_2 + k_{  {\rm{T}} } + {\overline \xi}_2 p_1 \;\; , 
\end{equation}
the forward kinematics  implies~\cite{jhep09}  
$(p_4+ p_6)^2  \gg  (p_3 +p_4)^2   $, $k_1 \simeq   \xi_1  p_1$, 
$k_2 \simeq    \xi_2  p_2 + k_{{\rm{T}} }$, so that 
\begin{equation}
\label{fwdkin}
p_5 \simeq  (1 - \xi_1 ) p_1 \;\;\;   ,  \;\;\;\;\;  p_6 \simeq   (1 - \xi_2 ) p_2 - k_\perp   
 \;\;\;   ,  \;\;\;\;\;  
\xi_1 \gg \xi_2     \;\;  .  
\end{equation}
In~\cite{jhep09} the full set  of the
 short-distance  matrix elements in the forward region,  needed 
for the evaluation of the k$_{\rm{T}}$-factorized 
   jet cross section (\ref{forwsigma}),  is computed 
in exclusive form, for all partonic channels.  

As discussed in~\cite{jhep09},    
these matrix elements,   though not    on shell,   are 
 gauge invariant and   perturbatively calculable. 
These  matrix elements are useful because   in the high-energy limit 
they factorize not only in the collinear emission region but also in the large-angle 
emission region. 
As long as 
 the factorization is carried out  in terms of distributions for parton splitting 
  at  fixed transverse momentum, 
they   can  serve to take  
into    account    effects of coherence from multi-gluon emission,  away 
from small angles,  
which become important for correlations among jets 
  across long      separations in rapidity.     We will  exploit this 
 in  performing  numerical  calculations for  central + forward jets    
in    Sec.~\ref{sec:f-jet-lhc}.

\subsection{Hard  matrix elements and merging }
\label{sec:matrel}

\begin{figure}[htb]
\vspace{60mm}
\includegraphics{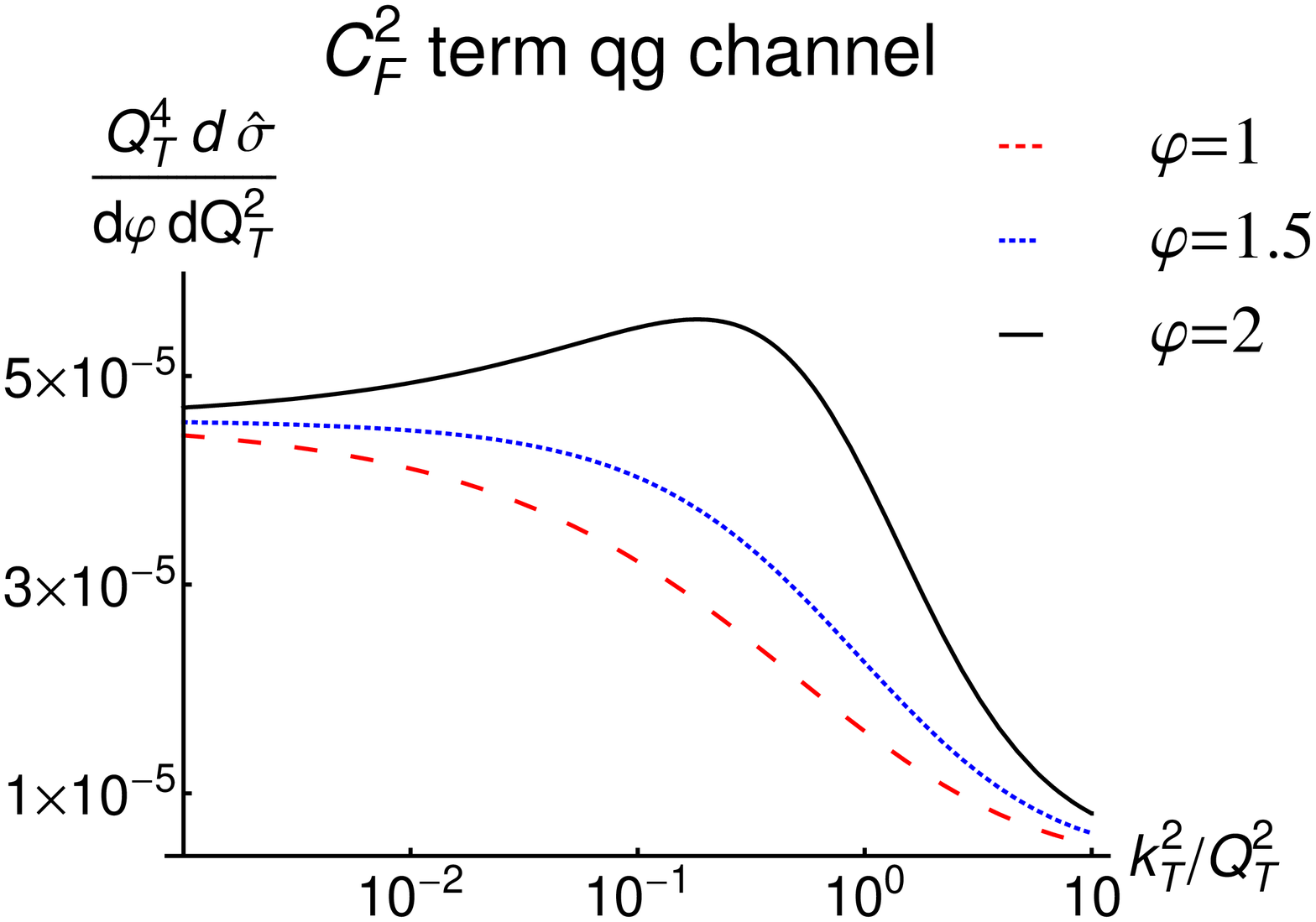}
\includegraphics{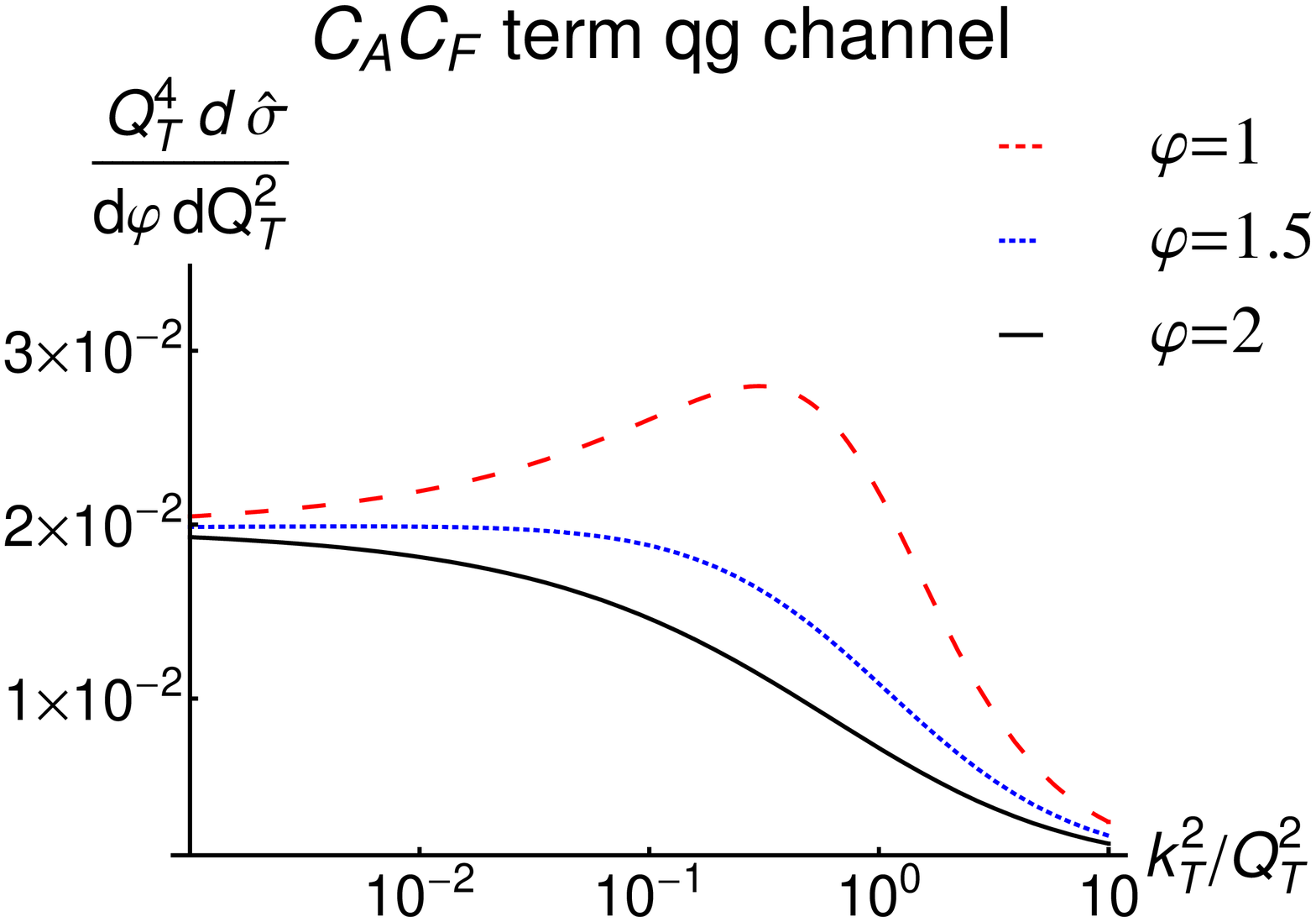}
\caption{\it Transverse momentum dependence of  the   factorizing  
short-distance  matrix elements. } 
\label{fig:ktbeh}
\end{figure}

The precise  behavior   in transverse momentum  resulting 
   from the  finite-angle radiation    taken into account by the method described  above 
   is illustrated in Fig.~\ref{fig:ktbeh}~\cite{jhep09}.  Here we consider  
    the $q g$ channel contribution to forward jet production.    
   (Analogous results for all channels can  be found in~\cite{jhep09}.) 
   We show separately the abelian  and non-abelian    terms,  proportional 
   respectively to the color factors $C_F^2$ and $C_A C_F$  ($C_F = (N_c^2 - 1) / (2 N_c)$, 
   $C_A = N_c$). 
  With reference to  Fig.~\ref{fig:sec2graph}, 
the variables  Q$_{\rm{T}}$ and $\varphi$   are 
 the final-state transverse momentum and azimuthal angle defined by  
\begin{equation}
\label{qtdef}
Q_T = (1-\nu) p_{T  4} - \nu p_{T  3}  \;\;, \;\;\;    {\rm{where}}  \;\;\; 
\nu =  (p_2 \, p_4) / [(p_2 \, p_1) -  (p_2 \, p_5)]   \;\;,  
\end{equation}
\begin{equation}
\label{phidef}
\cos \varphi = Q_T \cdot k_T / |  Q_T |  | k_T |   \;\; ,    
\end{equation}
  while   k$_{\rm{T}}$ is the transverse momentum defined by     
    Eq.~(\ref{kinek1k2}), and 
effectively     measures    
  the   distribution   
  of   the      jet  system    recoiling against the leading di-jets.

The region  k$_{\rm{T}}$   $ /  $   Q$_{\rm{T}}$  $ \to 0$  
in Fig.~\ref{fig:ktbeh} 
corresponds to 
the leading-order   process,  with two back-to-back 
jets.   The result in this region is simply given by the 
small-angle limit 
\begin{equation}
\label{born}
{{ Q_T^4 d   {\widehat  \sigma}   } \over 
{ d Q_T^2 d \varphi  }}  \to   \alpha_s^2 f^{(0)} ( p^2_T / s ) \;\; , \;\;\; 
Q_T \to p_T = | p_{T  3} | =  | p_{T  4} |      \;\; ,
\end{equation}
where the function $ f^{(0)}$  is given by~\cite{jhep09}    
\begin{equation}
\label{f0}
 f^{(0)} (z) = 
{1 \over  
{ 16  \sqrt {1- 4 z} } } \  \left[ C_F^2 z ( 1+ z ) + 2 C_F C_A ( 1 - 3 z  + 
z^2  ) \right]    \;\; . 
\end{equation} 

The summation of logarithms for large rapidity 
$y \sim  \ln s / p_T^2$, on the other hand, is  achieved 
   by  convolution   of the 
   k$_{\rm{T}}$ cross section     in  Fig.~\ref{fig:ktbeh}   with 
 unintegrated  parton-splitting  functions~\cite{hef}.  
While  in  the collinear approximation case  the  small-angle result in 
 Eqs.~(\ref{born}),(\ref{f0})   is  taken   to be valid, in leading order, 
  throughout the range of 
   transverse  momentum  scales up to  the factorization scale, 
 we see from 
 the result   in  Fig.~\ref{fig:ktbeh}  that 
  the role of  higher-order,  multi-gluon emission at large rapidities  
  is to   provide a modulation by 
  setting  the  dynamical cut-off     at values of k$_{\rm{T}}$   
   of order  Q$_{\rm{T}}$. The essential point is that 
  non-negligible 
  effects  may arise at high energy    from the finite-k$_{\rm{T}}$  tail.  
The quantitative evaluation  of these effects is obtained by  
 integrating   the  distribution     in 
 Fig.~\ref{fig:ktbeh} 
over the initial-state   parton  showers, also taken to be 
 transverse-momentum dependent (see  Sec.~\ref{sec:u-quark}).     
We will perform such  parton shower calculations explicitly in the 
next sections. 

Observe that,   as in any 
parton shower calculation beyond leading order, 
 in order to  combine the hard radiation encoded in the 
short distance matrix element with the radiation from parton showering  
one needs   
a specific  scheme for merging the two contributions 
consistently, by avoiding  double 
counting. The high-energy factorization of Subsec.~\ref{sec:fact}  
can be viewed as providing precisely such a merging scheme. 
In particular, we recall from~\cite{hef,cch-heraproc}
that the convolution of  off-shell  matrix elements over 
transverse-momentum dependent 
parton-splitting  functions  
is carried  out   by using systematically the 
relation  for the 2 + 1 jet cross section 
\begin{equation}
\label{distrib}
\int     \  d^2 k_T \    \left( {1 \over  k_T^2 } \right)_+    \  
  {\widehat  \sigma}  
(  k_T ) =   \int        \  d^2 k_T  \    {1 \over  k_T^2 } \ 
\left[   {\widehat  \sigma}  
(  k_T )  -   \Theta (\mu -  k_T ) \  {\widehat  \sigma}  
(  0_T ) \right]  
  \;\; ,  
\end{equation}  
which provides the necessary small-k$_{\rm{T}}$ subtraction.

So 
in Secs.~\ref{sec:u-quark}    and \ref{sec:f-jet-lhc} 
we will couple Eq.~(\ref{forwsigma})  to  parton showers  
and perform   parton shower calculations  {\it in the 
high-energy merging scheme}  for 
  hadronic 
final states associated with forward jets. 
 These  calculations  will 
illustrate    quantitatively  the  significance of  
contributions  
with   k$_T \simeq  Q_T$ in the large-$y$   region, and will   be  
compared with results of collinear-shower generators,  which do not 
include such   finite-k$_{\rm{T}}$  effects.   We note in particular  that the 
dependence on the azimuthal angle shown  in  Fig.~\ref{fig:ktbeh}
is  of  direct relevance, as we will see in Sec.~\ref{sec:f-jet-lhc},   for   
forward-region measurements   involving   azimuthal plane correlations 
between jets far apart in rapidity.

It is worth   noting that  the approach  summarized above 
 allows  forward    jets  to  be 
produced  either   from the hard  
scatter subprocess or from the parton evolution subprocess.    
We will see  an explicit numerical  illustration of  this  in Sec.~\ref{sec:f-jet-lhc}.  
This    picture 
 can  be contrasted with the 
 picture  from   collinear~\cite{michel_font} and BFKL~\cite{wallon10} approaches,  in which   
   forward jets   are    produced by   hard matrix elements or impact factors.   
This feature of the present  approach  
   can  be  traced back to  the fact   that the  factorization (\ref{forwsigma})  provides the 
   correct interpolation at high energy  between 
  the collinear  emission and   finite angle regions~\cite{hef}.

Because forward jet  production probes  the gluon  density function  
for small x  (see discussion around  Fig.~\ref{fig:sec2}),   
 it   can   naturally be used 
 to investigate  possible   
 nonlinear  effects~\cite{ianmue,hatta,gelis_etal_rvw} 
 at  high parton density.  We do not pursue  the study of such effects in the present paper;   
 but  we stress that 
 the formulation~\cite{jhep09} at fixed transverse momentum   
 is suitable to describe the approach to the high-density region, since, as  
 explained above, 
 it is designed to take  
  into account  both the  effects from  BFKL  evolution associated with the 
  increase in rapidity  and  also the   
 effects from   increasing  p$_T$  
 described by  renormalization group, which are  found to be also quantitatively 
 significant~\cite{kov10,weigert08,gardi06}  
  for studies of  parton saturation.  
We point the reader to  
e.g.~\cite{kutak-absorpt,avsar-stasto,avsar-iancu} for  
 first    Monte Carlo  calculations along these lines, 
 and~\cite{gelis_etal_hefnucl} for extension 
 to nucleus-nucleus collisions.

\section{Shower evolution and the unintegrated quark density }
\label{sec:u-quark}

To obtain a  detailed  description of the  hadronic 
final states associated with forward jets, we need  a full parton-shower 
calculation.   This section describes 
 basic features of the showering algorithm that we  use for such calculations.

As noted  earlier,  the factorization formula 
 in    Sec.~\ref{secfrom}   can  be used  jointly with  
  parton showering.  Because in the forward kinematics one of    the 
 longitudinal momentum fractions  x    in the  initial state 
    becomes small,      in order to take full  account of    
      multi-gluon emission    coherence  one needs to keep 
 finite-k$_{\rm{T}}$      terms     in the initial-state  parton 
      branching~\cite{hj_ang,mw92,Catani:1989sg,skewang}.\footnote{This brings in so-called 
      unintegrated, or transverse-momentum dependent, parton distributions. General 
      issues on these distributions are now actively investigated by many authors,  and  
      will  influence the use of 
        parton  branching methods. We comment on this in  
      Sec.~\ref{sec:conc}.  More  comments  may be found   in~\cite{hj_ang,hj_rec}.}  
 We will include these terms  according to the CCFM  
 method~\cite{cascade_docu,cascade_comp},  in 
 particular including also  showering 
 for 	quark  density channels  as  explained below.\footnote{Alternative methods 
 for  taking into account   finite-k$_{\rm{T}}$      terms in  the   parton 
 shower   are  considered  in~\cite{jadach09,watt_09,gustafson}.  
 See~\cite{hj_rec,acta09}    for  overviews   of the subject.}

The effect of  small-x     coherence terms    on   azimuthal 
 and transverse-momentum  jet correlations   and  jet  multiplicities   has been 
studied    in~\cite{hj_ang,hj_radcor}, focusing on the case  of  
 jet  lepto-production.  It is found that  quantitative effects  become more 
significant   with decreasing  x  and 
decreasing distance in  the  azimuthal plane   between the leading jets. 
The  results~\cite{hj_ang,hj_radcor}  suggest that   the inclusion of   small-x coherence 
  terms in the initial-state shower   can   be     relevant 
in the case of forward jets at the LHC.\footnote{Effects similar to 
those  computed   in~\cite{hj_ang} 
may also affect  azimuthal distributions 
of $b$ jets~\cite{phot09,azim_bjets}
and  jet multiplicities associated with Higgs boson production~\cite{deak_etal_higgs}. 
Early measurements of  forward jets at the LHC  can be helpful in this 
respect   to test how well  initial state radiation is  described  and/or 
for the QCD tuning of  Monte Carlo event 
generators. For the counterpart of this in the case of central jets see  the first 
LHC measurements~\cite{centr-jet-atl,centr-jet-cms}.} 

It was noted   by  
numerical  calculation   in~\cite{jhep09,proceed09}    that for 
realistic phenomenology  of  forward jets  in the LHC kinematics    
one needs to take into account contributions 
from  both  
quark-density and gluon-density channels.  
Since CCFM shower evolution 
 has typically only included 
gluon-density terms~\cite{cascade_comp}, here 
 we describe how we implement 
quark channels. 
For the  forward jet case that we are interested in, the quark density contributes 
at fairly large values of  x.   We will thus focus on the valence quark 
distribution.\footnote{Work to treat 
the sea  quark distribution  at unintegrated level is  
underway~\cite{jungetal_prep}.}   

We  consider the branching evolution equation   at the 
unintegrated, transverse-momentum dependent level  
 according to 
\begin{eqnarray}
x{ Q_v} (x,k_t,{\bar q} ) &=&  x{ Q_v}_0 (x,k_t,{\bar q} ) + \int \frac{dz }{z} 
\int \frac{d q^2}{ q^{2}} \ \Theta({\bar q} - zq) \ 
\nonumber\\ 
& \times&  \Delta_s  (   {\bar q} ,  zq)    
P  (z,  k_t) \  x{Q_v}\left(\frac{x}{z},k_t +  (1-z) q , q \right)     \;\; , 
\label{integral} 
\end{eqnarray}  
where   ${\bar q}$ is  the evolution  scale.  The quark splitting function  $P$ is  
given by 
\begin{eqnarray}
P   (z,  k_t)   &=& {\bar \alpha_s} \left(k_t^2 \right) 
 \frac{1+z^2}{1-z}    \;\;  , 
\label{splitt}
\end{eqnarray}
with $\alphasb=C_F \alpha_s  /  \pi $. 
Note that,  unlike the CCFM kernel  given in the appendix~B of~\cite{Catani:1989sg}, 
in Eqs.~(\ref{integral}),(\ref{splitt})    
 the non-Sudakov form factor is not included,  because  we only associate this 
  factor  to   $1/z$ terms. 
The Sudakov form factor $\Delta_s$ is given by   
\begin{equation}
\Delta_s(q_{i},z_iq_{i-1}) =\exp{\left(
 - \int_{z_{i-1}^2 q_{i-1}^2} ^{q^{2}_{i}}
 \frac{d q^{2}}{q^{2}} 
 \int_0^{1-Q_0/q}  \  \frac{1}{1-z}  \  \alphasb(q^2(1-z)^2)   \  dz
  \right)}         \;\; .   
  \label{Sudakov}
\end{equation}
 Here the fractional energy of the exchanged quark
$i$ is given by $x_i$,  and the energy transfer between the exchanged
quarks $i-1$ and $i$ is given by $z_i=x_i/x_{i-1}$.   The term $x{ Q_v}_0$ in 
    Eq.~(\ref{integral})  is the contribution of the non-resolvable branchings 
between  starting scale $q_0$ and evolution scale ${\bar q}$, given by 
\begin{equation}
 x{ Q_v}_0 (x,k_t,{\bar q} )   =    x{ Q_v}_0 (x,k_t,q_0 )  \Delta_s  (   {\bar q} ,  q_0)   \;\; ,   
  \label{Q0term}
\end{equation}
where $ \Delta_s$ is the Sudakov form factor, and the starting distributions  at scale $q_0$  are 
parameterized  using  the CTEQ5 
    $u$ and $d$ valence quark 
distributions~\cite{Lai:1999wy} as 
\begin{equation}
    x{ Q_v}_0 (x,k_t,q_0 ) =    x{ Q_v}_{\rm{CTEQ}} (x,q_0 ) \ 
    \exp[ - k_t^2 / \lambda^2 ]   \;\; ,   
  \label{gauss}
\end{equation}
with $\lambda = 0.92\; GeV$.

\begin{figure}[htbp]
\begin{center}
\epsfig{file=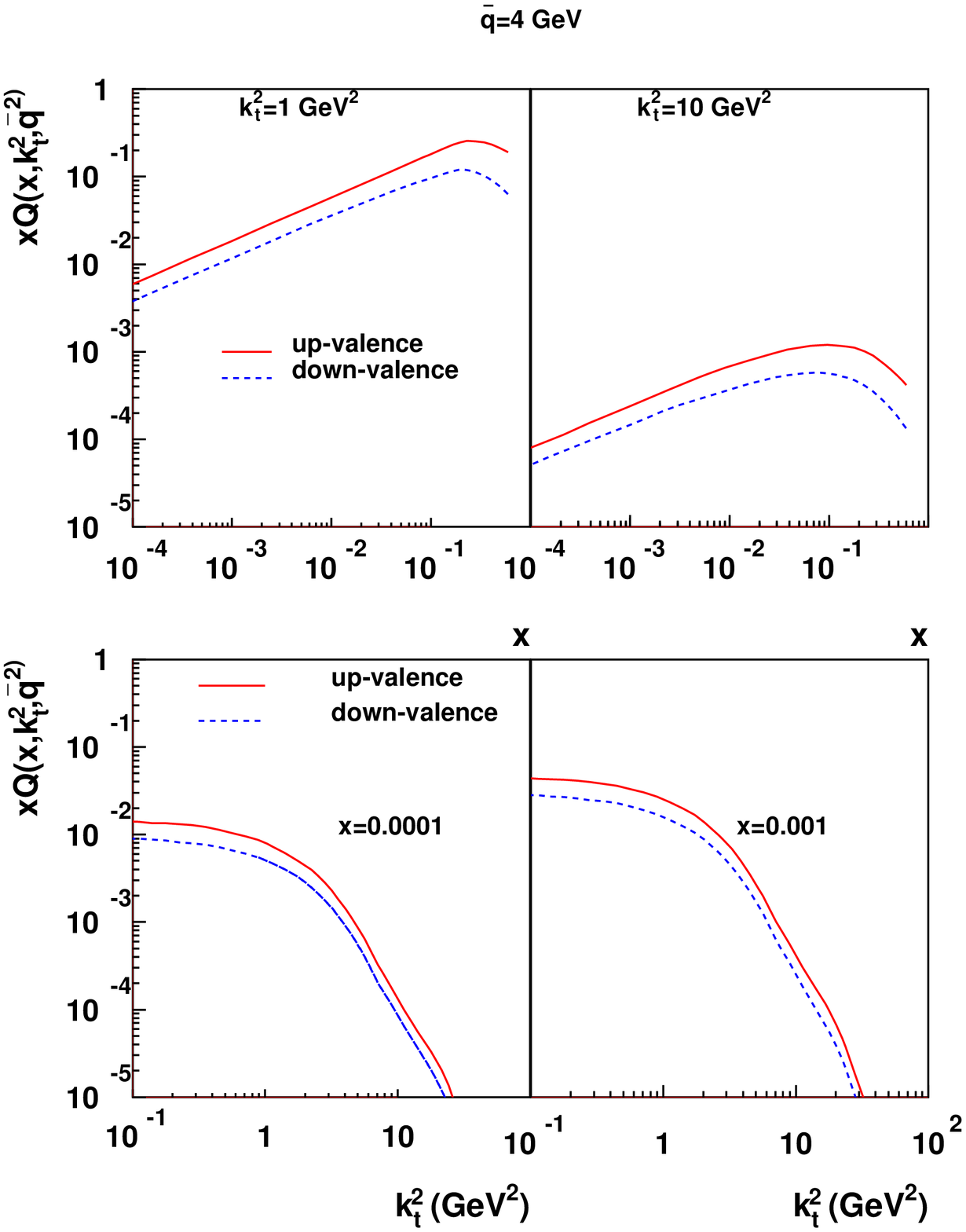,scale=0.4} \hskip -1cm
\epsfig{file=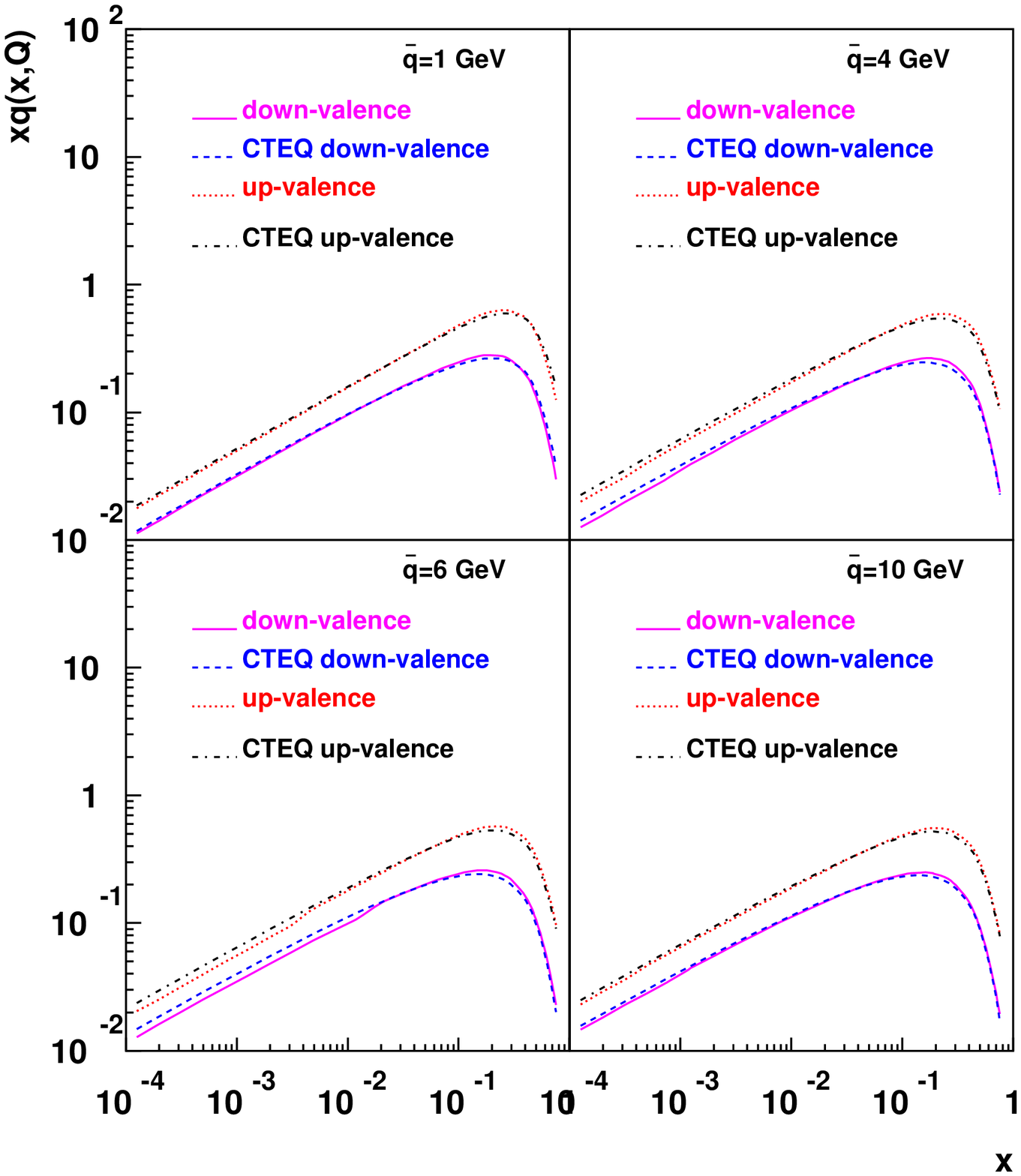,scale=0.4} 
\caption{\it {Left: Unintegrated quark distribution ($u,d$-quarks) as a function of $x$ at fixed $k_t$ (top) 
and as a function of $k_t$ at fixed $x$ (bottom) 
at a scale ${\bar q} = 4$ GeV.  Right: Integral of the unintegrated quark distribution ($u,d$-quarks) as a function of $x$  for different scales ${\bar q}$. Also shown is the $u,d$-quark distribution obtained from CTEQ~\protect\cite{Lai:1999wy}. }}
\label{Fig:uquark}
\end{center}
\end{figure}

We next  solve  Eq.~(\ref{integral})   numerically.
In fig.~\ref{Fig:uquark}(left) the unintegrated u-quark  and d-quark distributions   are  shown 
as a function of $x$ and as a function of $k_t$. In fig.~\ref{Fig:uquark}(right)  
we show the following 
integral of the quark distribution  
\begin{equation}
\int_0^{\bar q} x{ Q_v} (x,k_t,{\bar q} ) \ dk_t        \;\; ,  
  \label{integ-quark}
\end{equation}
and compare this for consistency with the distribution obtained    
from CTEQ~\cite{Lai:1999wy} 
 at the same scale.    

\section{Central + forward jet production at the LHC }
\label{sec:f-jet-lhc}

\begin{figure}[htb]
\vspace{85mm}
\includegraphics{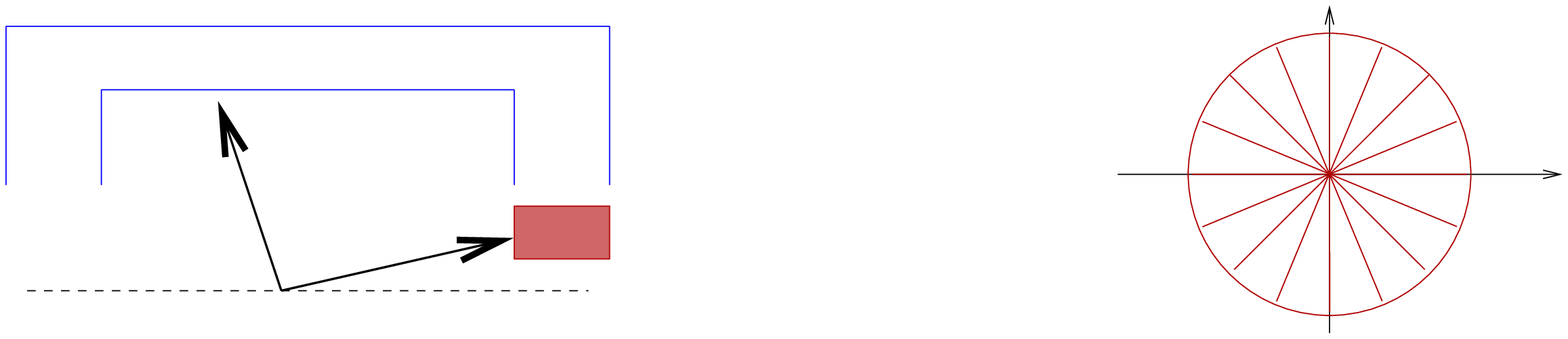}
\includegraphics{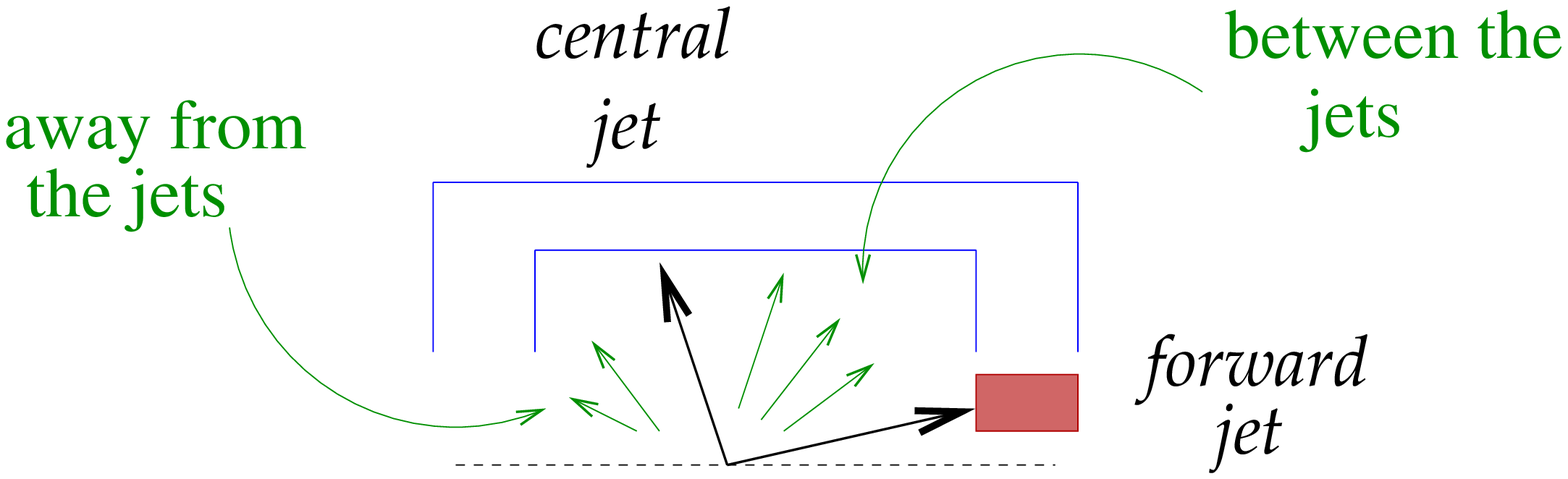}  
\caption{\it (top) Jets in the forward and  central detectors, and 
azimuthal plane segmentation;   
 (bottom) particle and energy flow in the inter-jet and outside regions.} 
\label{fig:jetcorr} 
\end{figure}

In a typical LHC experiment jets can be measured for high  
transverse energy 
E$_\perp\!>\! 30$~GeV in a large range of  pseudorapidity $\eta$. 
In the following we consider  differential cross sections  
 for dijets   (Fig.~\ref{fig:jetcorr})    
 reconstructed with the  Siscone algorithm~\cite{fastjetpack}   with $R = 0.4$,          
 where one jet is  in the central region defined by $|\eta_c|\!<2 $ and the other jet is in the forward region defined by $3\!<\!|\eta_f|\!<5$. 

\subsection{Dijet cross section and $\Delta R$ distribution }
\label{sec:cent_forw}

\begin{figure}[t!]
\vspace{5.5cm}
  \begin{picture}(30,0)
    \put(40, -40){
      \includegraphics{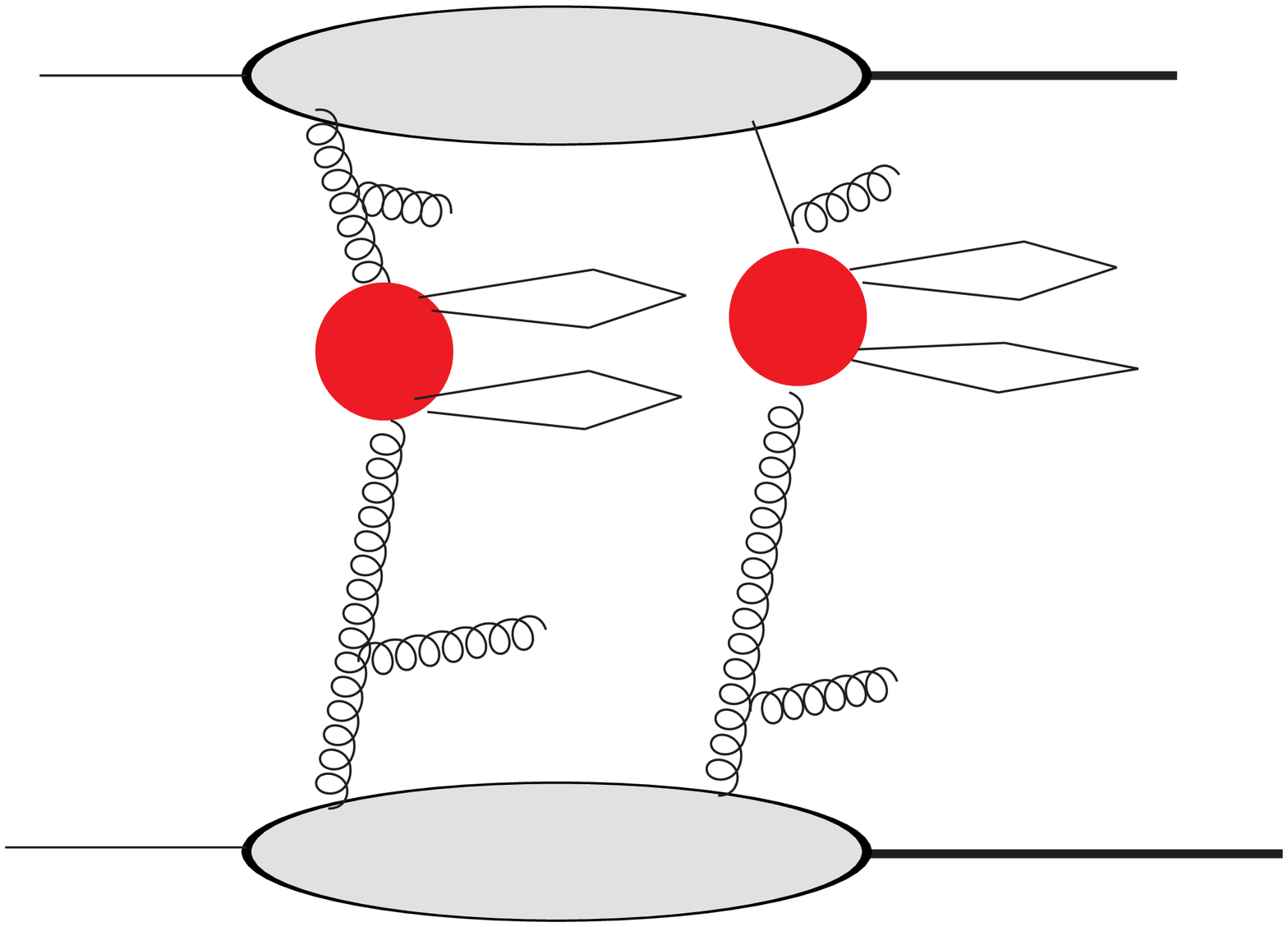}
    }
    \put(270, -40){
      \includegraphics{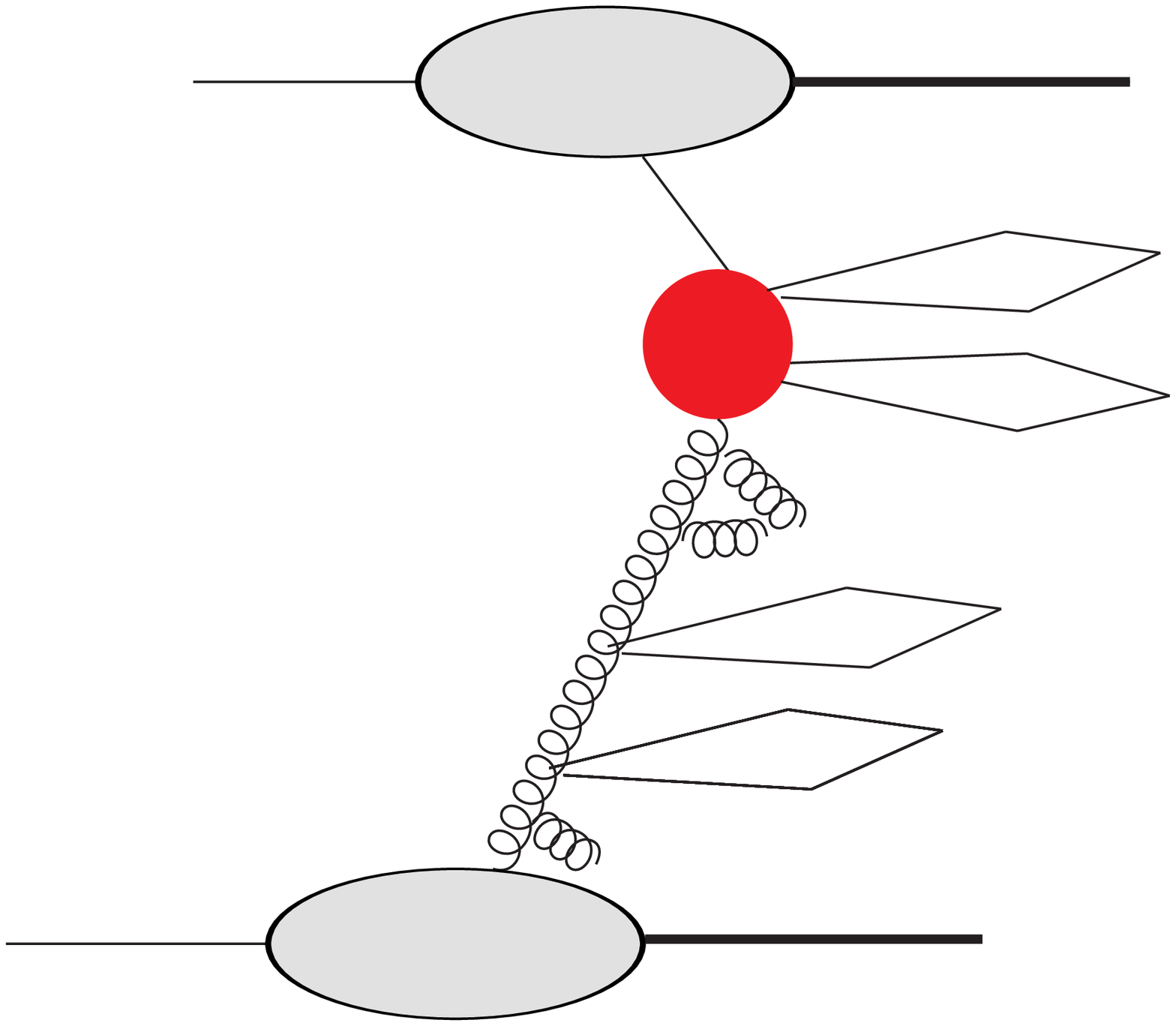}
    }    

     \end{picture}
\caption{\it Multi-jet production by 
 (left) multiple  parton chains; (right) single parton chain.} 
\label{fig:mpi}      
\end{figure}

The total cross section    for    a   central and a  forward jet    
obtained with the \CASCADE\ ~\cite{cascade_docu} Monte Carlo event 
generator (version 2.2.03 
 including the matrix element calculated in~\cite{jhep09}, the unintegrated gluon distribution {\bf set A} and the unintegrated valence 
quark distribution described in section~\ref{sec:u-quark}) is given in Tab.~\ref{xsection}.
We compare the prediction from \CASCADE\ with the prediction from the  \PYTHIA~\cite{Sjostrand:2006za} Monte Carlo event generator running in two  modes: with and without multi-parton interactions 
 (Fig.~\ref{fig:mpi}).  
We use tune P1\cite{pz_perugia}, which allows for more radiation from parton shower. 
Both Monte Carlo generators  simulate higher order QCD corrections with parton showers: \CASCADE\ uses parton showers according to the CCFM evolution equation whereas \PYTHIA\ uses DGLAP based parton showers. 
\begin{table}[htdp]
\caption{\it Integrated dijet cross section for $E_T > 10 (30) $~GeV in the range
$|\eta_c|\!<2 $ and $3\!<\!|\eta_f|\!<5$.}
\begin{center}
\begin{tabular}{|l|r|r|}
\hline
  &$\sigma (E_T>10\;GeV)$&$\sigma (E_T>30\;GeV)$\\ \hline 
 \protect\CASCADE\ & 469  $\mu$b  & 3.1 $\mu$b  \\ \hline 
 \protect\PYTHIA\  (MPI)  (P1)   &  798 $\mu$b & 3.5 $\mu$b \\ \hline 
 \protect\PYTHIA\  (no MPI)    & 346 $\mu$b &  3.3 $\mu$b \\ \hline 
 \end{tabular}
\end{center}
\label{xsection}
\end{table}%
The total cross section predicted by \CASCADE\  lies in between the prediction of 
\PYTHIA\ with and without multiparton interactions (Tab.~\ref{xsection}).
\begin{figure}[htbp]
\begin{center}
\epsfig{file=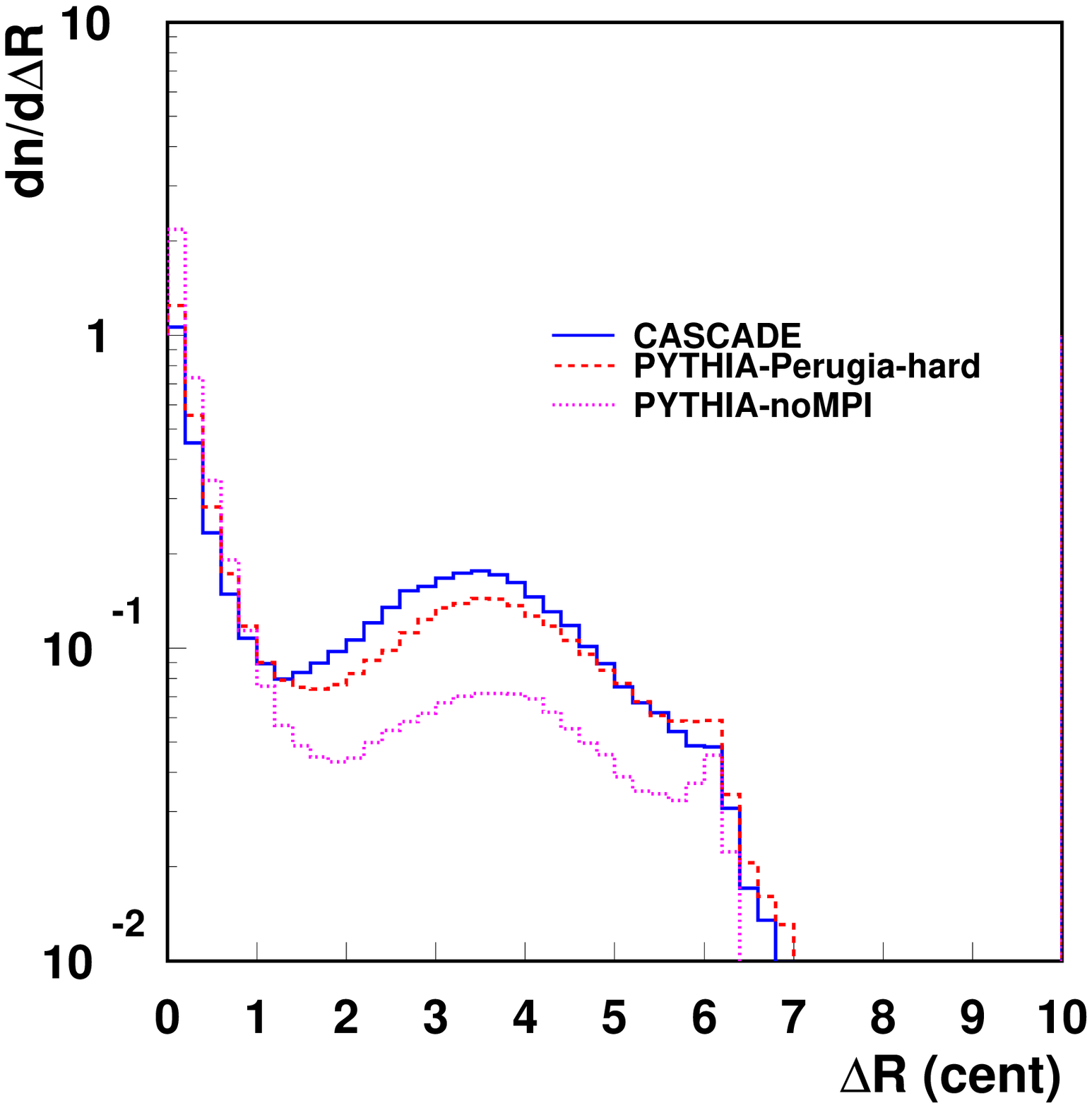,scale=0.4} \epsfig{file=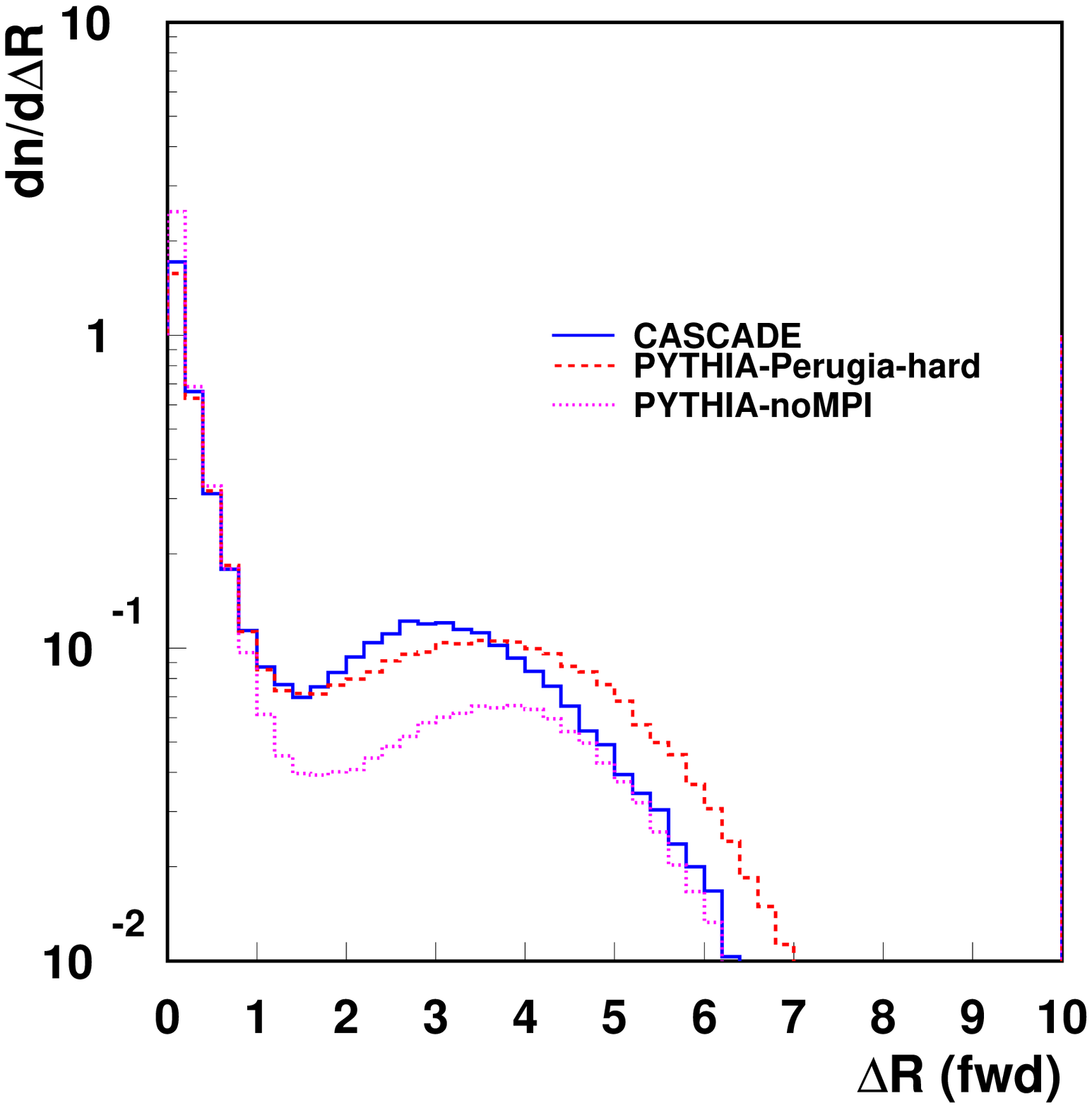,scale=0.4} 
\epsfig{file=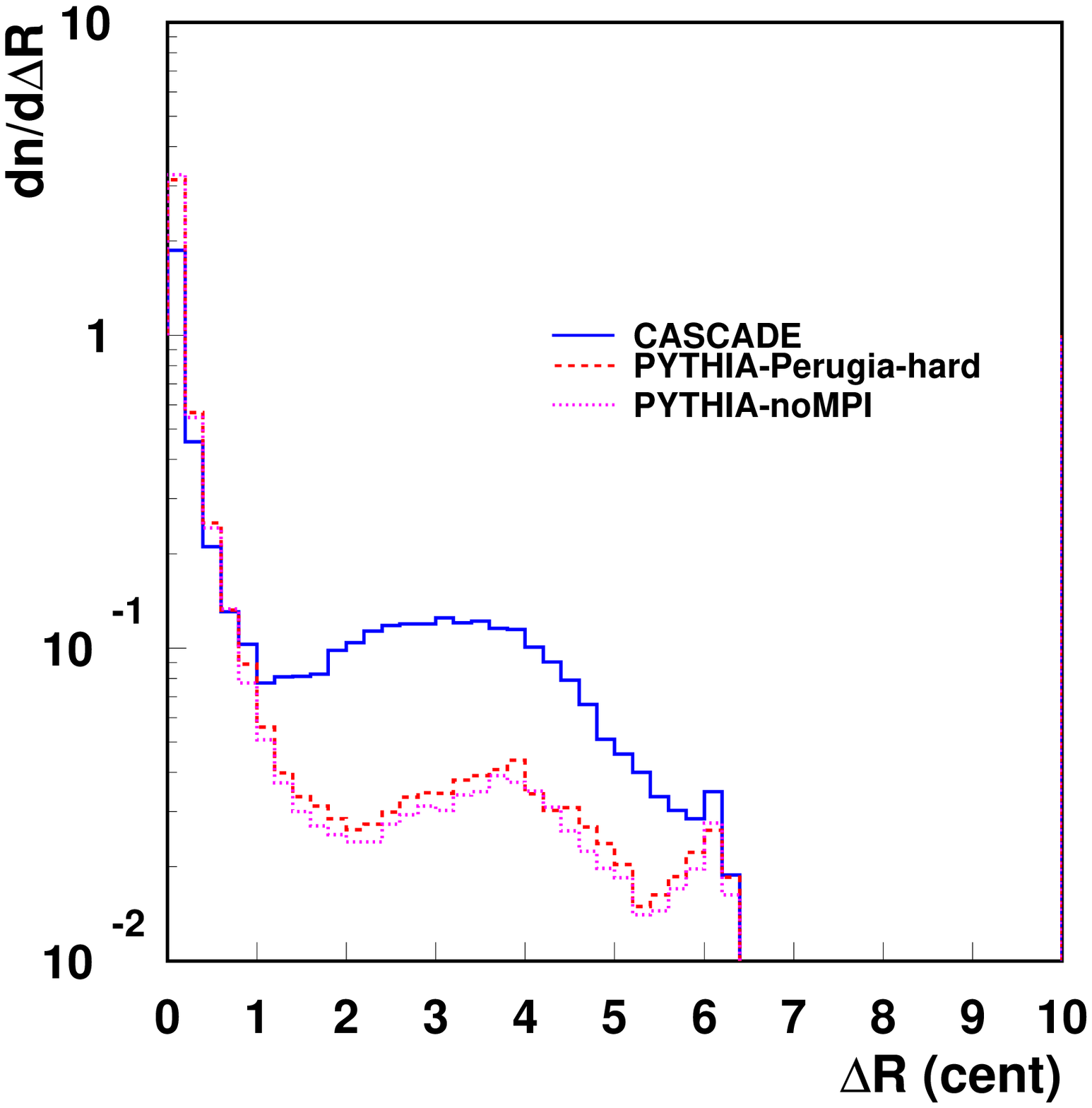,scale=0.4} \epsfig{file=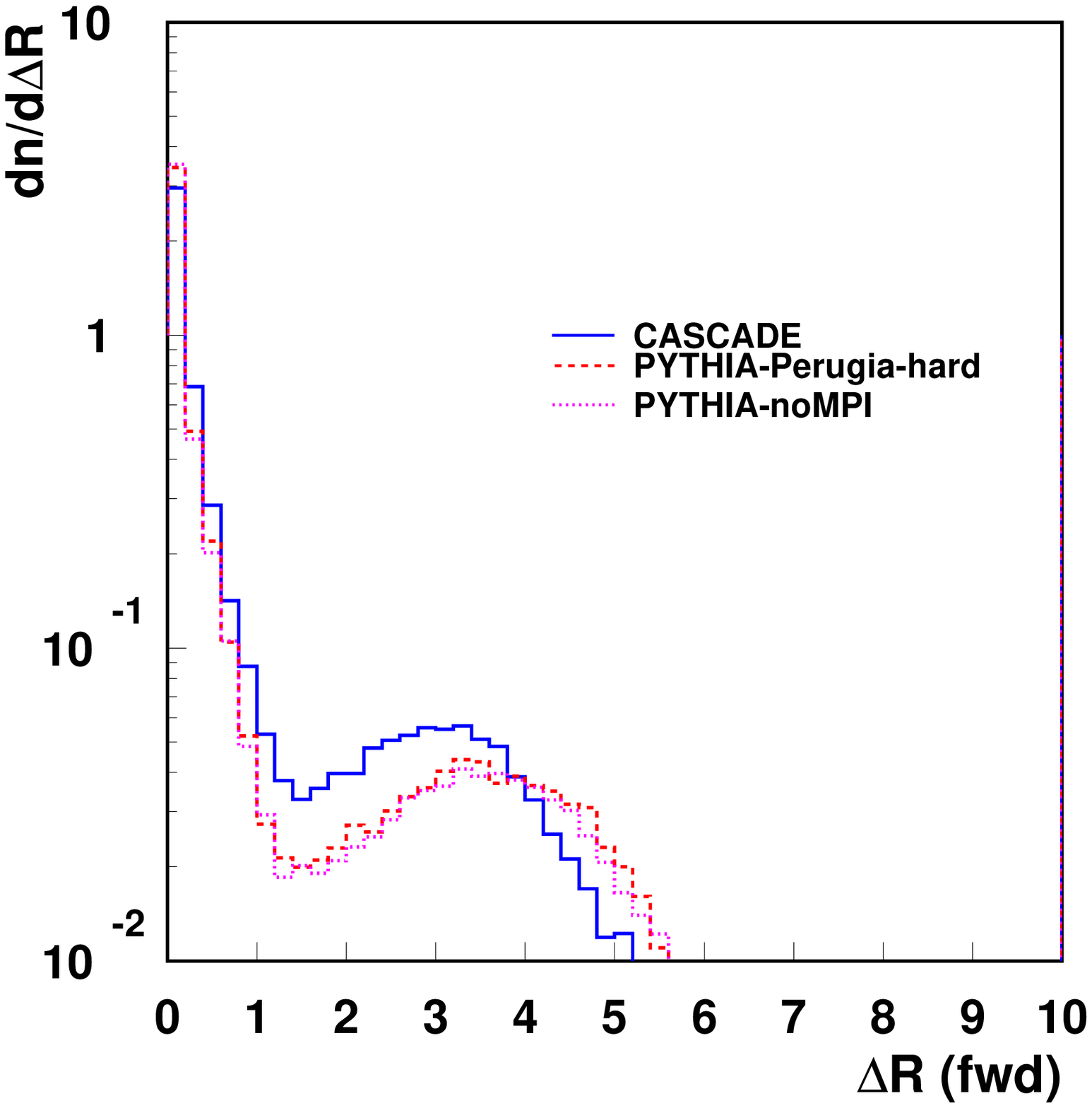,scale=0.4} 
\caption{\it {$\Delta R$ 
distribution of the central  ($|\eta_c|<2$, left)  and forward jets ( $3 < |\eta_f| <5 $, right) 
 for $E_T > 10$~GeV (upper row) and  $E_T>\!30$~GeV (lower row). 
The prediction from the k$_\perp$ shower (\protect\CASCADE) is shown with the solid blue  line; 
 the prediction from the collinear shower (\protect\PYTHIA)   including  multiple interactions  and without multiple interactions  is shown with the red and purple lines.   }}
\label{Fig:DeltaR}
\end{center}
\end{figure}

In   Fig.~\ref{Fig:DeltaR}    
   we investigate the  physical mechanism  producing  the central and forward jets.  
 We plot $\Delta  R = \sqrt{(\Delta \phi)^2 + (\Delta \eta)^2 }$, where $\Delta \phi= \phi_{jet} - \phi_{part}$ ($\Delta \eta= \eta_{jet} - \eta_{part}$) is the azimuthal (pseudorapidity) difference between the jet and the corresponding parton from the matrix element. The    $\Delta  R$  distribution 
  allows  one 
 to see 
  whether the jets are dominated by hard partons from the matrix element, or whether they originate from the parton shower. 
For the low $E_T$ jets the distribution in $\Delta  R$ has a significant contribution from jets not corresponding to a parton from the matrix element ($\Delta R > 1$). 

The bump 
 structure of the distribution at $\Delta R >1 $ is consistent with random distributions in 
 $\Delta \phi$ and $\Delta \eta$ within the phase space region investigated here, showing that indeed $\Delta R >1$ corresponds to the region where there is no correlation between the parton from the matrix element and the jet. 

It is interesting to observe  that \CASCADE\ predicts a similar distribution as obtained from \PYTHIA\ with multiparton interactions, whereas \PYTHIA\ without multiparton interactions has a significantly smaller contribution at large $\Delta R$.
The situation changes for the high $E_T$ jets: \CASCADE\ predicts significantly more jets (especially in the central region) coming from the parton shower as compared to \PYTHIA . This behavior is understandable, as the 
small-x initial-state 
 parton shower allows for higher transverse momentum radiation compared to a collinear parton shower. It also shows, that the most forward jet is mainly coming from the matrix element parton, whereas the central jet  has a significant contribution from the parton shower.

\subsection{Transverse momentum spectra}
\label{sec:trv}

\begin{figure}[htbp]
\begin{center}
\epsfig{file=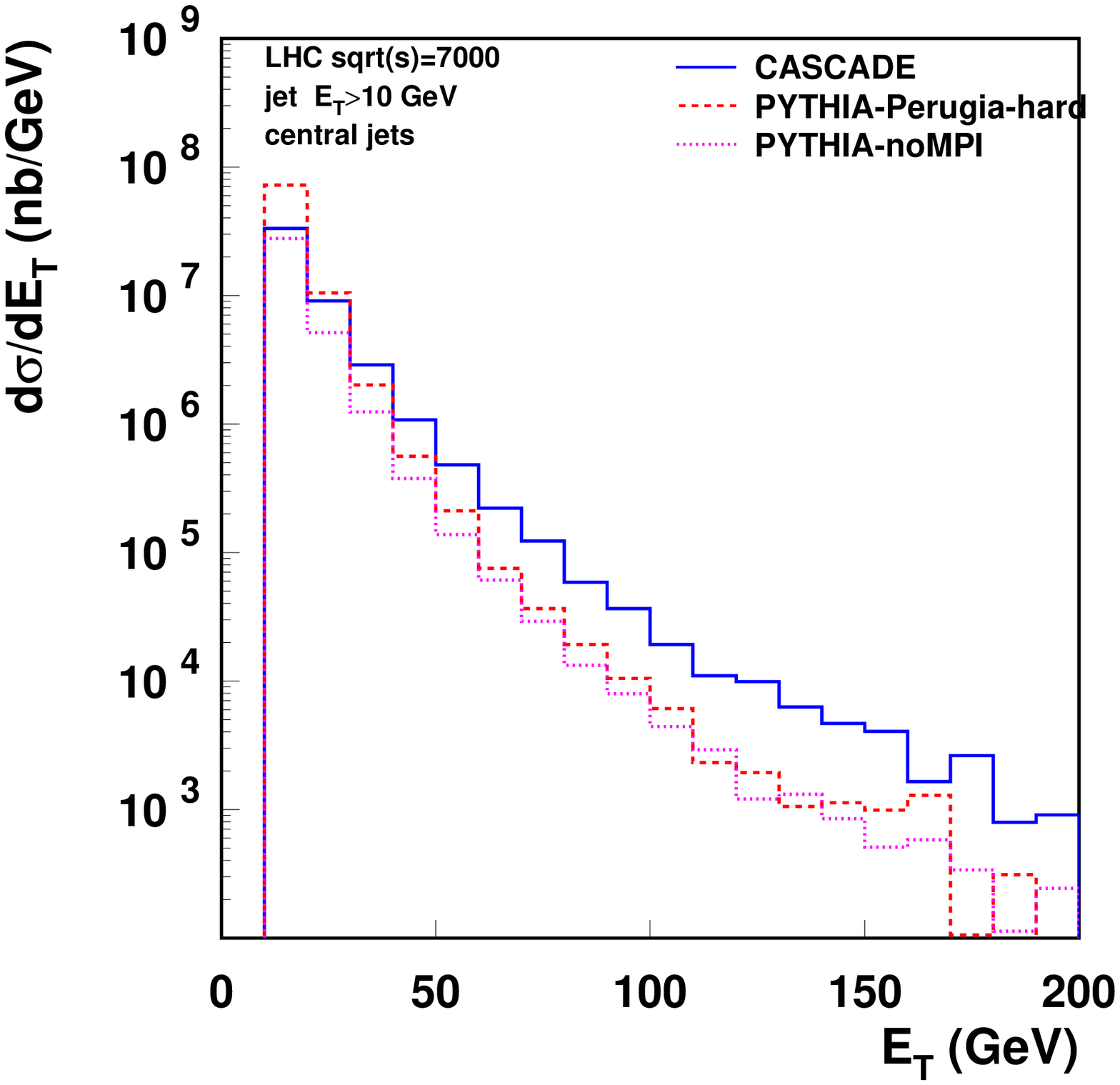,scale=0.4} \epsfig{file=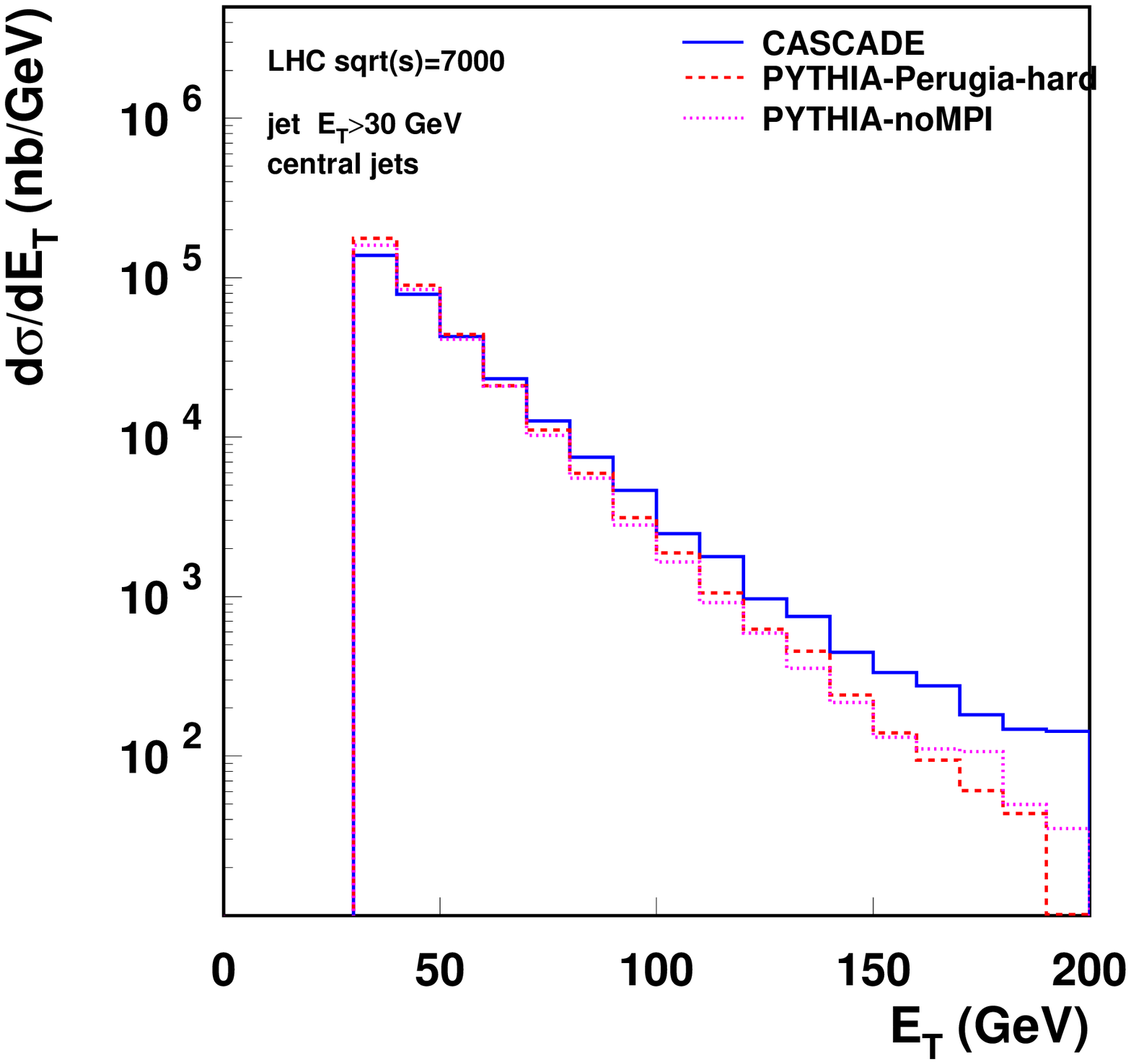,scale=0.4} 
\caption{\it {Transverse momentum spectra of central jets with $|\eta_c|<2$  at $\sqrt s=7$~TeV for $E_T > 10$~GeV (left) and for $E_T>\!30$~GeV (right) for events which have a forward jet with $E_T>10 (30)$~GeV in $3 < |\eta_f| <5 $.
The prediction from the k$_\perp$ shower (\protect\CASCADE) is shown with the solid blue  line; 
 the prediction from the collinear shower (\protect\PYTHIA)   including  multiple interactions  and without multiple interactions  is shown with the red and purple lines.  }}
\label{Fig:transversal_central}
\end{center}
\end{figure}

\begin{figure}[htbp]
\begin{center}
\epsfig{file=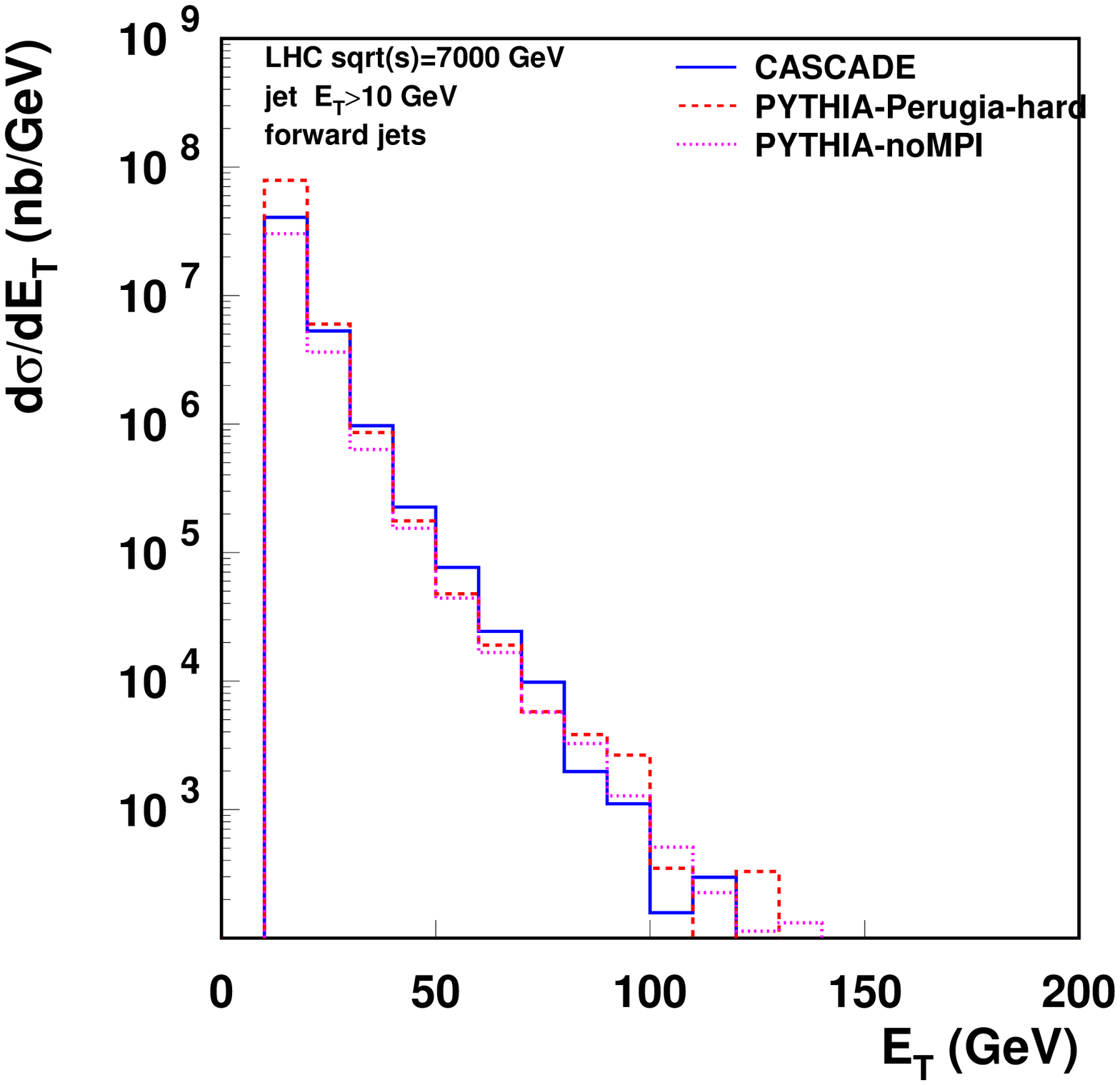,scale=0.4} \epsfig{file=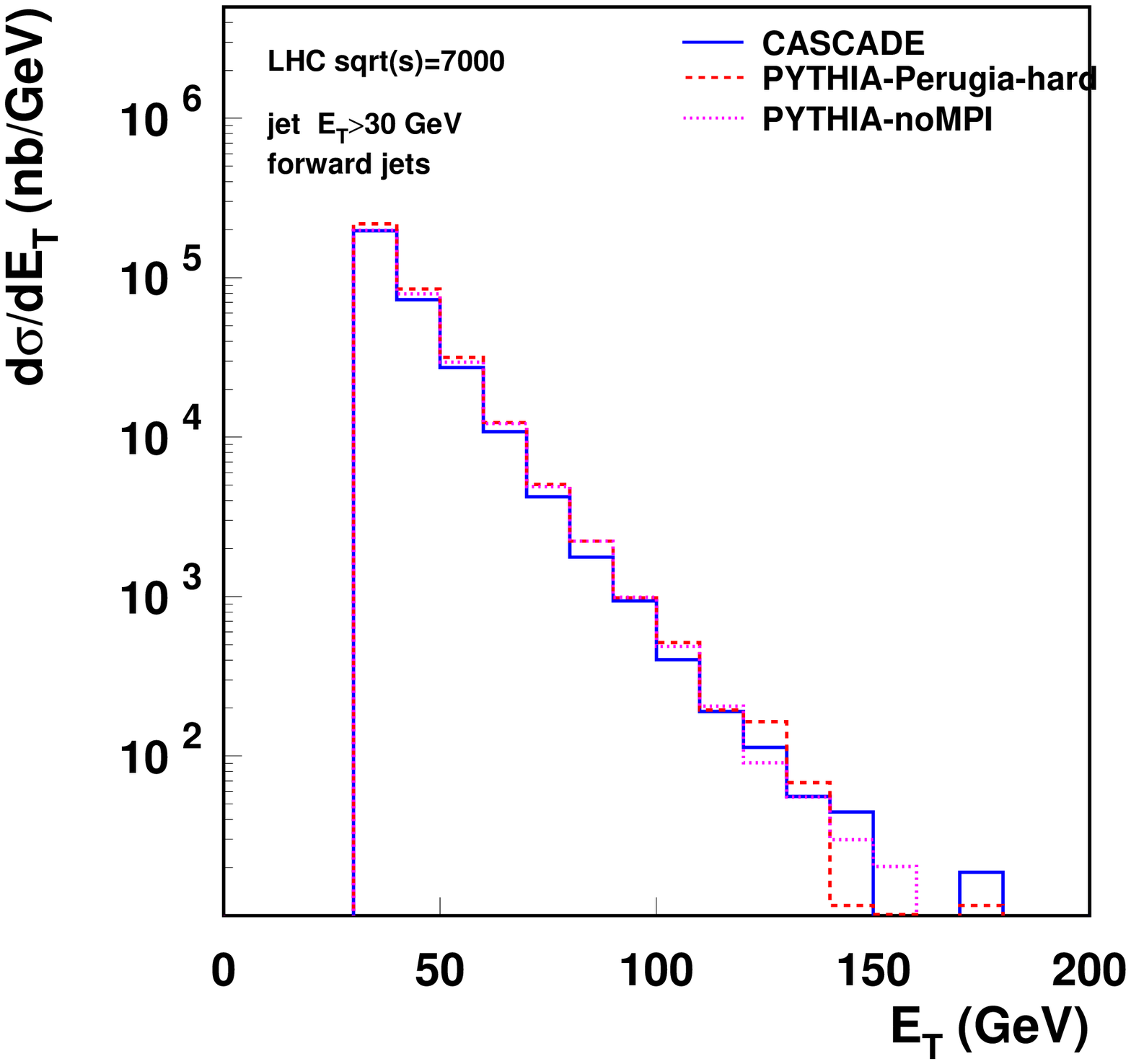,scale=0.4} 
\caption{\it {Transverse momentum spectra of forward jets with $3<|\eta_f|<5$  at $\sqrt s=7$~TeV for $E_T>\!10$~GeV (left) and for $E_T>\!30$~GeV (right) for events which have a forward jet with $p_\perp\!>\!10 (30)$~GeV in $ |\eta_c| <2 $.     
The prediction from the k$_\perp$ shower (\protect\CASCADE) is shown with the solid blue  line; 
 the prediction from the collinear shower (\protect\PYTHIA)   including  multiple interactions  and without multiple interactions  is shown with the red and purple lines. }}
\label{Fig:transversal_forward}
\end{center}
\end{figure}

 Figs.~\ref{Fig:transversal_central} and \ref{Fig:transversal_forward}   show 
  the differential cross section 
${d\sigma} / {dE_T}$ for jets reconstructed with the 
Siscone 
algorithm in the central region defined by $|\eta_c|\!<2 $ and in 
   the forward region defined by $3\!<\!|\eta_f|\!<5$.  
The left(right) plots show the results   when  both jets have $E_T > 10 (30)$~GeV.
We see that the  k$_\perp$-dependent 
 parton shower  
  implemented in \CASCADE\  produces a significantly harder spectrum especially for the jet in the central region when both jets are required to have $E_T > 10$~GeV. The predictions from \PYTHIA\ with and without multiparton interactions are similar at large $E_T$, whereas the multiparton interactions contribute significantly in the low $E_T$ region ($E_T < 30$~GeV) resulting in a larger cross section.

This behavior can be understood 
 since  \CASCADE\ uses matrix elements which are calculated within  high-energy factorization,   allowing  harder transverse   
momentum dependence as compared to collinear factorization. Moreover \CASCADE\ 
uses the CCFM parton shower with angular ordering which at small  
x allows for a random walk in transverse momentum, 
and thus allows for more and harder parton radiation compared to  DGLAP based parton shower as implemented in \PYTHIA . 

It is also interesting to observe  that the contribution from multiparton interactions as implemented in 
 \PYTHIA\ 
is important only for the  region of low $E_T$ jets. 
We have checked also 
 the prediction obtained from  
 \PYTHIA\     
 using a different parameter set  for the  modeling  of multiple-parton chains 
 (tune D6T~\cite{rdf1}).   We find that  
 the difference between the different tunes for 
 the multiparton interaction parameters is smaller than the difference coming from the  noncollinear corrections 
 to single-chain parton shower.   

To see better   the significance of the results in  Figs.~\ref{Fig:transversal_central} and \ref{Fig:transversal_forward}, 
we have also considered  
 the next-to-leading-order Monte Carlo generator \powheg~\cite{alioli}. We find that  the 
 $E_T$ spectra from \powheg ,   in the range of rapidity and transverse energy considered here,  
 are very close to those of \PYTHIA , indicating that the 
 enhancement  from the k$_\perp$ shower in 
 Figs.~\ref{Fig:transversal_central} and \ref{Fig:transversal_forward}
 is not  simply due to the next-to-leading term, but comes from corrections beyond 
 next-to-leading order.    This feature is  noteworthy; 
  its study should be extended to  a wider range of forward-jet observables.

\subsection{Rapidity dependence}
\label{sec:rap}

\begin{figure}[htbp]
\begin{center}
\epsfig{file=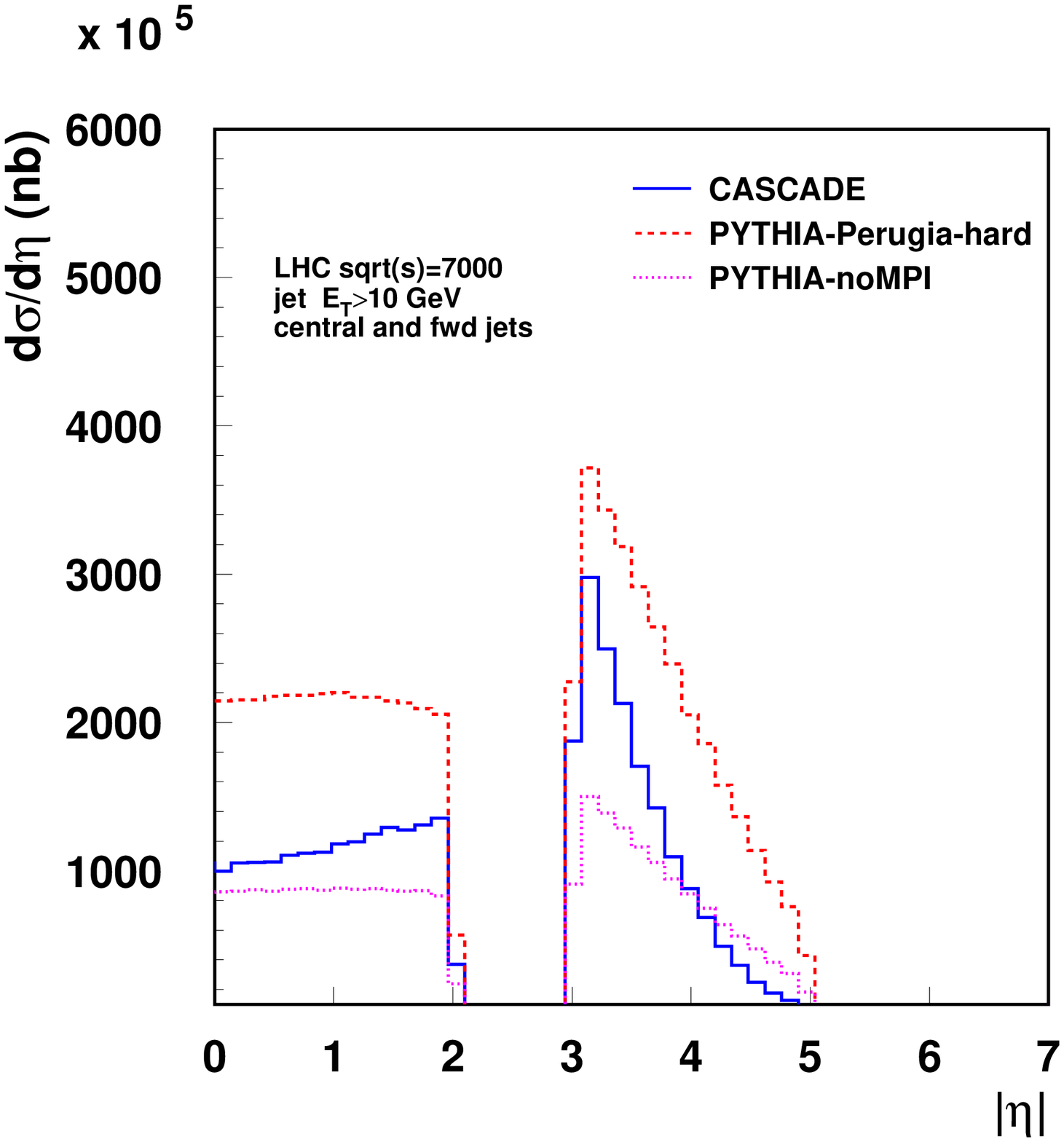,scale=0.4} \epsfig{file=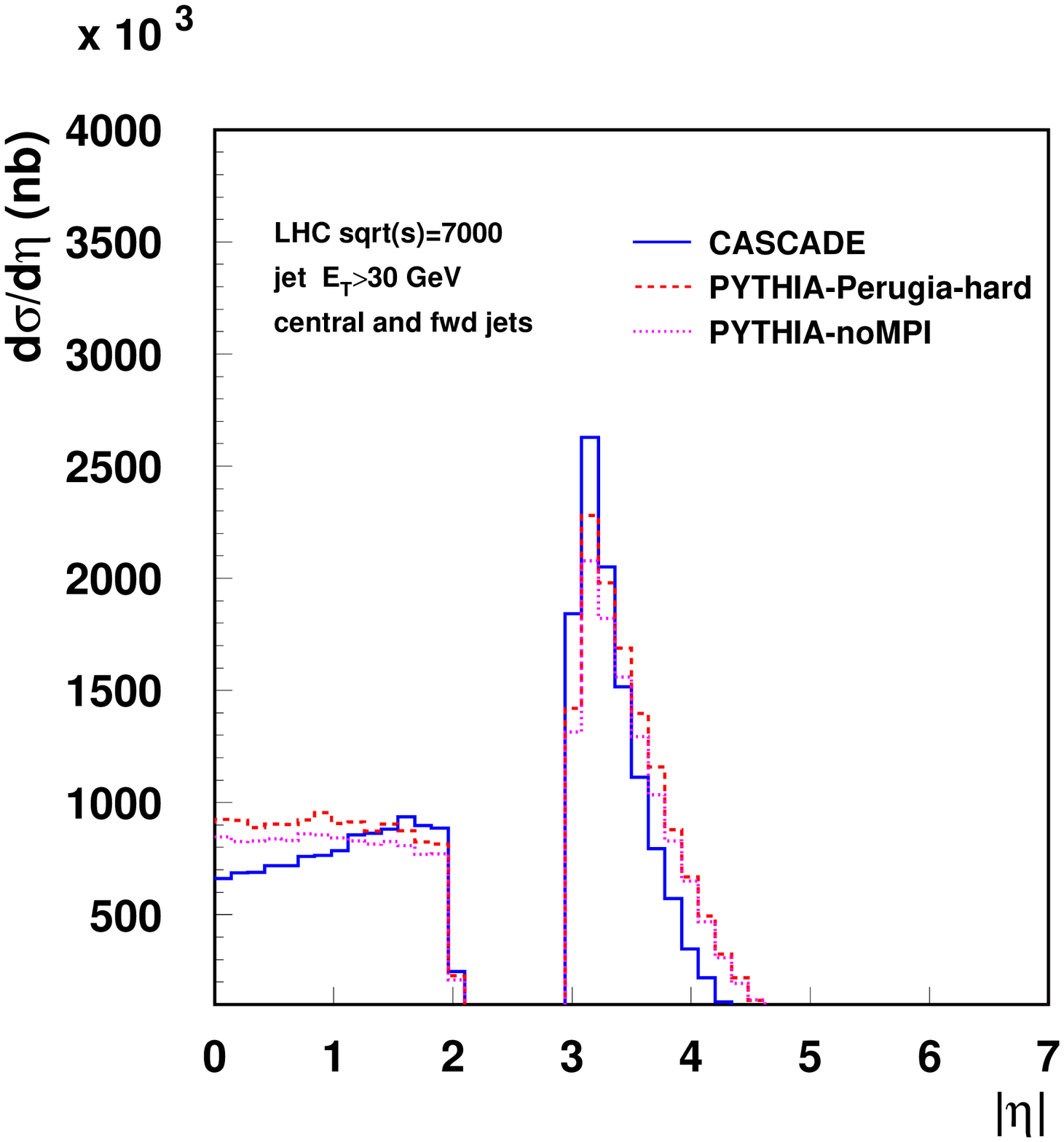,scale=0.4} 
\caption{\it {Pseudorapidity spectra of produced jets for  $\sqrt s=7\,TeV$ with requirement that $p_T\!>\!10~GeV$(left) and  $p_T\!>\!30~GeV$ (right). 
\protect\CASCADE\ is shown with the blue solid line,  \protect\PYTHIA\ with (without)  multiparton interactions is shown as the dashed red (dotted purple) line.
 }}
\label{Fig:rapidity}
\end{center}
\end{figure}

In Fig.~\ref{Fig:rapidity} we show the differential cross section ${d\sigma} / {d\eta}$  for dijet events with $E_T > 10 (30)$ GeV in two regions of  $0\!<\!|\eta|\!<\!2$ and $3\!<\!|\eta|\!<\!5$. Again we compare the prediction from \CASCADE\ with the one from \PYTHIA\ without and with multiparton interactions.

We observe that the cross section in the central region for \CASCADE\ is rising towards larger $\eta$ whereas for \PYTHIA\ the cross section is flat. 
The cross section in the forward region is steeply falling towards large $\eta$. The slope of this distribution is different from \CASCADE\ and \PYTHIA .   A closer 
 investigation of these different behaviors 
is underway. A significant  contribution to  the difference comes from 
 the treatment  of the 
quark distribution (Sec.~3), and suggests   the need 
to include both valence and sea quark 
distributions  at unintegrated level.

From  Fig.~\ref{Fig:rapidity} (left)  we see that at small $E_T$ the influence from multiparton interactions in \PYTHIA\ is significant. \CASCADE\ predicts a cross section of similar size as \PYTHIA\ with multiparton interactions in the region $3 < \eta <3.5$, but the distribution falls more rapidly towards larger $\eta$.  In the large $E_T$ region (Fig.~\ref{Fig:rapidity} (right)) 
the differences in cross sections as a function of $\eta$ become smaller.

\subsection{Azimuthal dependence}
\label{sec:phi}

\begin{figure}[htbp]
\begin{center}
\epsfig{file=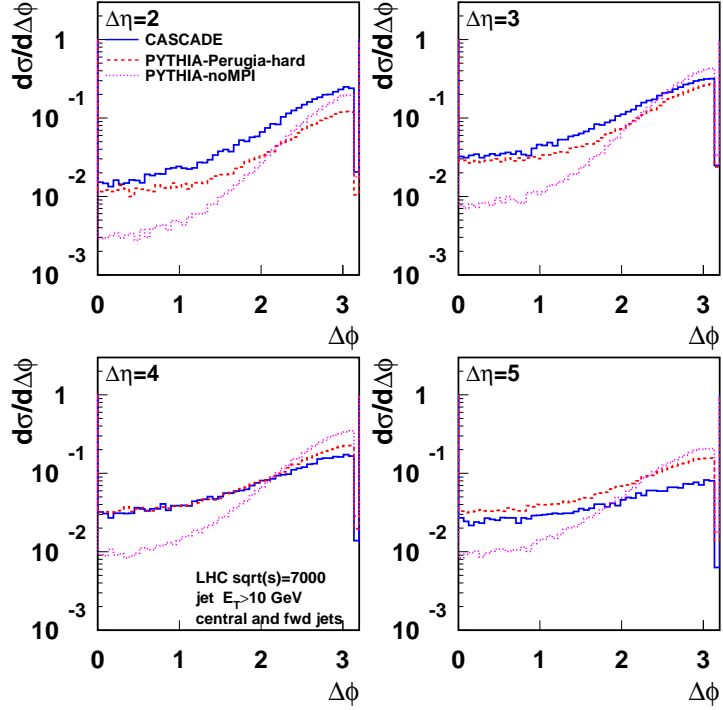,scale=0.44} \epsfig{file=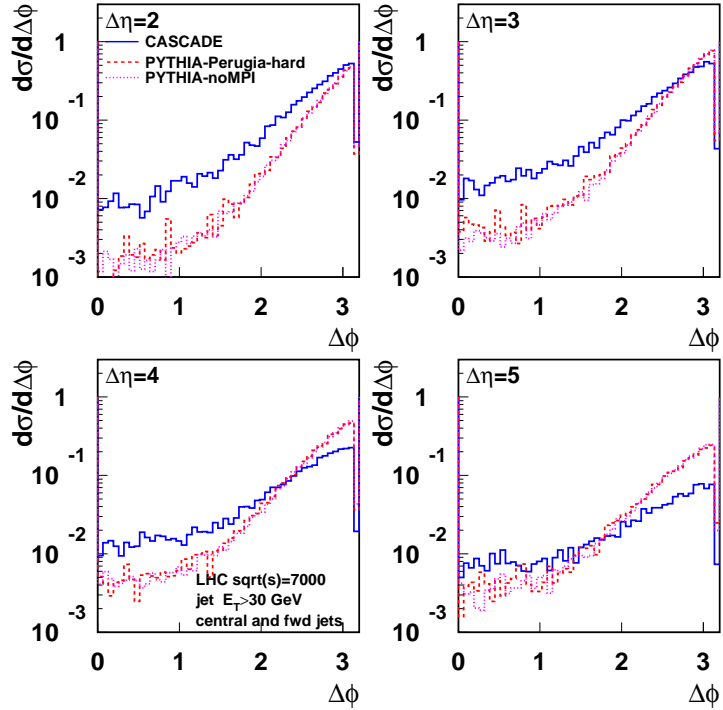,scale=0.44} 
\caption{\it {Cross section as a function of the azimuthal difference $\Delta \phi$  between the central and the forward jet for different  separations in pseudorapidity $\Delta \eta$, 
 at  $\sqrt s=7\,TeV$ for jets with $p_T\!>\!10~GeV$(upper) and  $p_T\!>\!30~GeV$ (lower). 
\protect\CASCADE\ is shown with the blue solid line,  \protect\PYTHIA\ with (without)  multiparton interactions is shown as the dashed red (dotted purple) line.
 }}
\label{Fig:deltaphi}
\end{center}
\end{figure}

The azimuthal correlation of a central and forward jet is a measure of the 
parton radiation  between the jets and is therefore a probe of  how well 
the Monte Carlo  parton shower  is    simulating 
 the higher-order parton emissions. The azimuthal decorrelation of forward and backward jets has been proposed as one of the measurements to test BFKL 
 dynamics~\cite{wallon10}(and references therein). In lowest order, BFKL predicts a much larger decorrelation compared to calculations in collinear factorisation. With a larger separation 
 of the jets in $\Delta\eta$, the phase space for parton radiation is increased. 
However, significant multiparton interactions could perhaps mimic a signal expected from small x dynamics.  

In Fig.~\ref{Fig:deltaphi} we show the differential cross section ${d^2 \sigma}  / 
{d\Delta\phi \Delta\eta}$. The decorrelation as a function of $\Delta\eta$ increases in \CASCADE\ as well as in \PYTHIA . 
In the low $E_T$ region (Fig.~\ref{Fig:deltaphi} (left)) the increase in decorrelation with increasing $\Delta\eta$ is very significant. The cross section for jet separation up to $\Delta\eta < 4 $ is very similar between \CASCADE\ and \PYTHIA\  with multiparton interactions, whereas a clear difference is seen to \PYTHIA\ without multiparton interactions. However, at large $\Delta\eta > 4 $ the decorrelation predicted by \CASCADE\ is significantly larger than the prediction including multiparton interactions. 

In the higher $E_T$ region \CASCADE\ predicts everywhere a larger decorrelation. In this region, the influence of multiparton interactions in \PYTHIA\ is small and the difference to \CASCADE\ comes entirely from the different parton shower. 
\begin{figure}[htbp]
\begin{center}
\epsfig{file=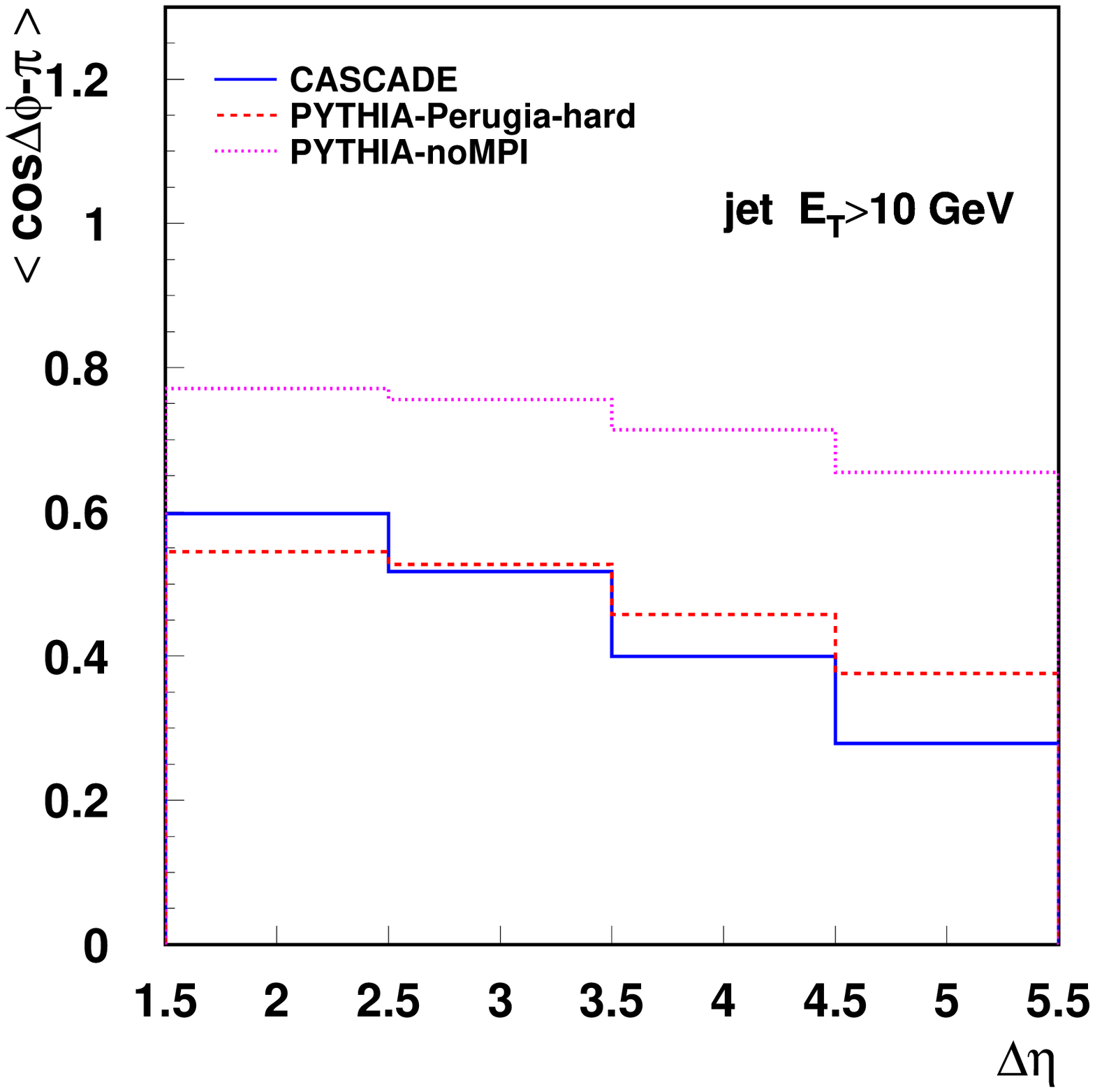,scale=0.4} \epsfig{file=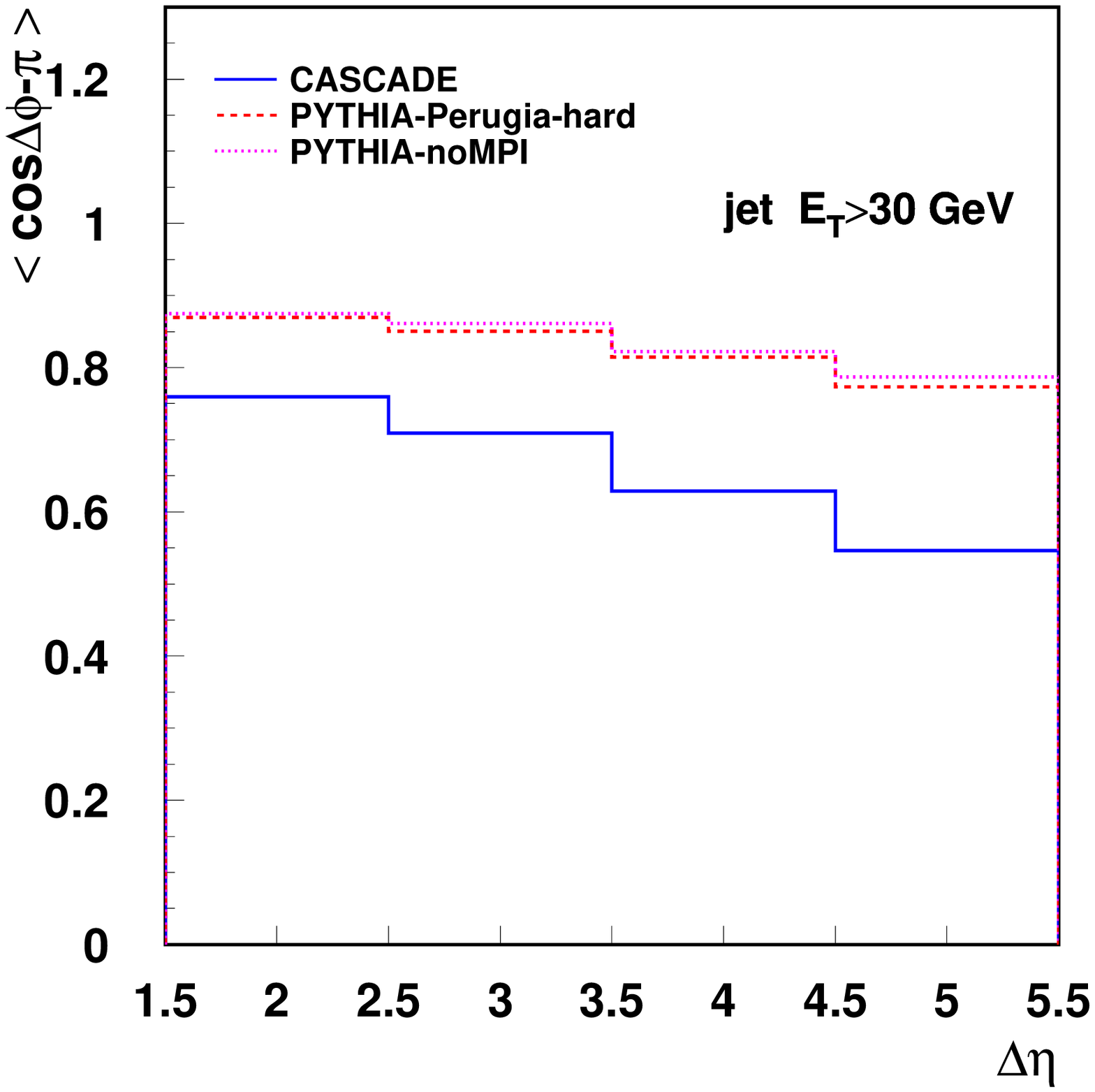,scale=0.4} 
\caption{\it {Average   $\langle \cos(\Delta \phi - \pi) \rangle$ between the central and the forward jet as a function of  separation in pseudorapidity $\Delta \eta$, 
 at  $\sqrt s=7\,TeV$ for jets with $p_T\!>\!10~GeV$(left) and  $p_T\!>\!30~GeV$ (right). 
\protect\CASCADE\ is shown with the blue solid line,  \protect\PYTHIA\ with (without)  multiparton interactions is shown as the dashed red (dotted purple) line.
 }}
\label{Fig:avdeltaphi}
\end{center}
\end{figure}

In Fig~\ref{Fig:avdeltaphi} we show the average 
$\langle \cos(\Delta \phi - \pi) \rangle$ as a function of the rapidity 
separation $\Delta \eta$ of the central and forward jets. This quantity is  considered 
in~\cite{michel_font,orrsti,mn_pheno,wallon10} as a sensitive probe for 
BFKL dynamics. We observe 
 that the distribution of $\langle \cos(\Delta \phi - \pi) \rangle$ at $p_T\!>\!10~GeV$ shows only  little difference between \CASCADE\ and \PYTHIA , whereas the differential distribution of ${d^2 \sigma}  / 
{d\Delta\phi \Delta\eta}$ (Fig.~\ref{Fig:deltaphi})  is  more discriminative. In the $p_T\!>\!30~GeV$  case \CASCADE\  predicts a  larger decorrelation than   \PYTHIA , consistent with what is 
observed in Fig.~\ref{Fig:deltaphi}.

\section{Summary and outlook}
\label{sec:conc}

The production of  hadronic  jets  in the forward region of pp collisions will form a largely new 
area of experimental and theoretical activity at the Large Hadron Collider.    
Forward jet production will enter the LHC physics program both for   new particle discovery 
processes (e.g., vector boson fusion channels for Higgs boson searches) and for new aspects of 
standard model physics (e.g., QCD at small x and its interplay with cosmic ray physics). 

In this paper we have focused on  the study of  jet correlations for production of a forward 
and a central jet at the LHC. The  capabilities  of  LHC  forward and central detectors 
allow  measurements of such correlations  to be made for the first time across 
large  rapidity intervals  ($\Delta y   \greatersim 4 \div 6$).    
In this kinematic region the evaluation of QCD theoretical predictions is made complex 
by the presence of multiple mass scales, which raises questions on the potential need for, on one 
hand,  
perturbative QCD resummations  and,  on the other hand,   
possible corrections beyond single parton scattering.

To address the structure of the final states associated with forward jet production, 
in this  paper  we have used a merging scheme based on 
high energy factorization~\cite{hef,jhep09} 
to combine hard  matrix elements and parton showering.  
This is designed  to incorporate  radiative contributions    
in such a manner   that  both 
logarithmic corrections   in  the large  rapidity  interval  
and logarithmic corrections  in  
the hard transverse momentum are taken into 
account  to higher orders in $\alpha_s$.     Both kinds of contributions are  likely to be   
needed   for  reliable  phenomenology of  forward hard production.  
Their  summation  is achieved  by   including 
 finite-k$_{\rm{T}}$  terms  in the parton branching and   matrix elements.

In this approach    
forward jets may be produced from either the hard scatter subprocess or  
the parton showering subprocess. 
The resulting physical picture of forward hard production  
  is thus  different from that 
of  purely BFKL or   collinear  calculations,  in which, also at  the 
next-to-leading order (see e.g.~\cite{michel_font,wallon10}),    
forward jets   are    produced by   hard  impact factors  or  matrix elements.

We have implemented the high-energy  merging scheme 
 in the hadron-level  shower Monte Carlo  event generator   
\CASCADE~\cite{cascade_docu},    and used this   to obtain   numerical  predictions 
for  several forward-jet observables.  
In doing this, we have also implemented in the   Monte Carlo generator~\cite{cascade_docu} 
an algorithm  to  include the parton branching evolution for the valence quark  
distribution at unintegrated level,  which extends the  previous  
CCFM algorithm~\cite{cascade_comp}  for  the unintegrated gluon distribution.

We have  examined  the effects  of   the   higher-order  radiative  contributions 
taken into account by this approach   
by computing  the jet transverse-momentum  and rapidity spectra and the jet 
correlations in rapidity and azimuth.  We find that  the effects are significant 
especially in the slope of the $E_T$ spectrum and in the jet angular  correlations.  
In particular, we find that while the average cosine 
 of the  azimuthal separation between the leading jets is not  affected very  much   as a function of 
 rapidity  by  finite-k$_{\rm{T}}$  terms, 
 the detailed shape of the  $\Delta \phi$ distribution is. 
 
 We have also  investigated  the model~\cite{pz_perugia,Sjostrand:2006za}       
 for   multiple parton interactions, corresponding  to  corrections   beyond single parton 
 scattering.   
 Our analysis shows that  certain features of  forward jet production 
 such as rapidity and $E_T$ spectra,    
 found by  including 
   high-energy, noncollinear corrections to single-chain parton showers,  
 can be mimicked      by  effects of  multiple  parton chains.    
However,  distinctive shapes   are found both 
 in the $\Delta R$ distribution and in the  azimuthal correlations. 
We suggest that measurements of the  particle and energy flow  in the region both between the jets 
and away from the jets  should have  stronger discriminating power  between the 
single-chain and multiple-chain mechanisms for multi-jet production.  The detailed 
  analysis of this point will be the subject of a separate  paper.   Note that this also 
   points to the  phenomenological relevance of   energy flow observables  such as those 
   investigated    in~\cite{bryan_eperp_10,sung,yatta-ueda}.  

As  observed in Secs.~\ref{secfrom} and \ref{sec:u-quark},     many of the 
  theoretical tools  that underlie  forward jet  physics, from  
   parton  branching beyond leading order to 
   perturbative QCD resummations to,  possibly,  the approach to  the saturation region,      
 depend on   the notion of transverse momentum dependent, or unintegrated, 
parton distribution functions (u-pdfs).   
In the calculations of  this  paper  we take the   high-energy 
 definition of u-pdfs~\cite{hef}, namely, 
we  rely on  the fact that  for small x 
  u-pdfs can be defined 
gauge-invariantly  (and  can be  
related to the ordinary pdfs renormalized 
in the minimal subtraction scheme ${\overline{\rm{MS}}}$~\cite{hef93})     
by going to  the high-energy pole in physical amplitudes~\cite{hef}.  
More general characterizations of u-pdfs,  valid over the whole phase space,  
are very desirable. A    complete   framework 
  is yet to be  fully developed though. 
Recent  results in  
   this area, see 
   e.g.~\cite{becher-neub,rogers,idilbisci,chered,bacch-diehl,rogers08,fhfeb07},  
    are likely  to     eventually   have a bearing    on forward jet physics.

Note that forward hard production processes will 
be relevant   not only for  LHC physics  but  also  for physics at 
 the planned future   lepton facilities~\cite{laycock} (LHeC, EIC). 
Thus a  unified understanding in hadron-hadron and lepton-hadron collisions is 
 desirable.   As recalled  in Sec.~\ref{secfrom},  
  QCD  high-energy factorization~\cite{hef} 
 has been used to determine   the 
 asymptotic coefficients~\cite{forwjetcoeff}  that couple 
  forward jets to  deeply inelastic scattering. 
Since the early phenomenological studies~\cite{forwdis92}, forward jet 
leptoproduction has been investigated at HERA.  Measurements of 
 forward jet cross sections at   HERA~\cite{heraforw} have illustrated that 
 neither fixed-order next-to-leading  
 calculations  nor  standard shower 
 Monte Carlo generators~\cite{webetal99,heraforw,knut}, e.g.   \pythia\ 
  or \herwig,  are   able to   describe 
 forward jet  ep data.  
 This   provides  additional  motivation for   developing    
 methods    capable of   treating the multi-scale 
  kinematics and 
   describing   jet   production    beyond  the central rapidity region. 
It is   of  interest to analyze HERA  
data~\cite{aaron-hera}   looking at  forward + central jets, 
similarly to what is  done in this paper    (Sec.~\ref{sec:f-jet-lhc}) for pp collisions; but  
 due to the  phase space available for  multiple jet radiation, 
 such studies  are likely to  prove much  more  relevant  at a future high-energy lepton collider.

The analysis  performed in this paper  shows that 
 the   final states associated with forward jet   production at the LHC 
  receives significant contributions 
 from   radiative corrections that take into account both 
 large logarithms of rapidity and 
 large logarithms of transverse momentum. Distinctive  effects are found,   for instance,  
for  the distribution in the azimuthal separation  $\Delta \phi$ 
 between  forward and  central jets.   We obtain 
 distinctive predictions both  with respect to  parton showers modeling 
 multiple parton interactions and with respect to parton showers including 
 next-to-leading fixed-order corrections.
This analysis can be extended to correlations of  forward and backward jets. It can thus serve 
to estimate the size of backgrounds  from  QCD radiation 
  in new particle searches from vector boson fusion channels.

\vskip 1 cm

\noindent    {\bf Acknowledgments}. 
We thank the Terascale Physics Analysis Center, DESY, the 
CERN Theory Division and  LHC Physics Center for hospitality 
at various stages during this work.  
M.~D.\  thanks   UniverseNet and IFT/CSIC for financial support.   
We acknowledge useful conversations with 
G.~Marchesini,  Z.~Nagy, J.~Proudfoot,  C.~Roda   and P.~Skands.

\bibliographystyle{mysty} 
\raggedright 
\bibliography{ref}

\providecommand{\href}[2]{#2}\begingroup\raggedright\begin{thebibliography}{1}

\bibitem{grothe}
        M.~Grothe, arXiv:0901.0998 [hep-ex]. 

\bibitem{ajaltouni}
         Z.~Ajaltouni et al.,   arXiv:0903.3861 [hep-ph].


\bibitem{denterria} 
       D.~d'Enterria, arXiv:0911.1273 [hep-ex].  


\bibitem{vbf-cms}
        M.~Vazquez Acosta  [on behalf of CMS Coll.],  arXiv:0901.3098 [hep-ex]. 
         
\bibitem{vbf-atlas} 
         K.J.C.~Leney   [on behalf of ATLAS Coll.],    arXiv:0810.3144 [hep-ex].       

   
\bibitem{albrow_review}
      M.G.~Albrow, T.D~Coughlin and J.R.~Forshaw, arXiv:1006.1289 [hep-ph].
      
   

\bibitem{cep_james_etal}
         L.A.~Harland-Lang et al.,          arXiv:1005.0695 [hep-ph].

\bibitem{fwdmssm}
       S.~Heinemeyer et al.,   
           arXiv:1009.2680 [hep-ph].

\bibitem{engel}
                R.~Engel,  in Proceedings of the 30th International Cosmic Ray Conference,  
              ed.~R.~Caballero et al., vol.~6, p.~359 (Mexico City,  2009). 

\bibitem{michel_font}              
              P.~Aurenche, R.~Basu and M.~Fontannaz, 
              Eur.\ Phys.\ J.\  C {\bf 57} (2008)   681; 
              M.~Fontannaz, LPT-Orsay preprint 09-86 (April 2009).    

 

\bibitem{muenav}
        A.H.~Mueller and H.~Navelet,     Nucl. Phys.  B{\bf 282} (1987) 727. 

\bibitem{webetal99}
             C.~Ewerz,     L.H.~Orr,  W.J.~Stirling   and   
             B.R.~Webber,  J.\  Phys.\ G{\bf 26} (2000) 696; 
             J.~Forshaw,   A.~Sabio Vera  and   
             B.R.~Webber,  J.\  Phys.\ G{\bf 25} (1999) 1511. 
            
            

\bibitem{orrsti}
       L.H.~Orr and W.J.~Stirling,    Phys.\ Lett.\  B{\bf  436}  (1998)  372.  



\bibitem{mn_pheno}
              A.~Sabio Vera        and F.~Schwennsen,       Nucl. Phys.  B{\bf 776}  (2007) 170;
              C.~Marquet and C.~Royon,       Phys.\  Rev.\  D{\bf 79} (2009) 034028. 



\bibitem{wallon10} 
       D.~Colferai, F.~Schwennsen, L.~Szymanowski and S.~Wallon, 
     JHEP {\bf 1012} (2010) 026;       arXiv:1010.0160 [hep-ph]. 
 
 
      
       
\bibitem{hef} 
     S.~Catani, M.~Ciafaloni and F.~Hautmann,    
     Phys.  Lett.  B{\bf 242}  (1990) 97;        Nucl. Phys.  B{\bf 366} (1991) 135.  
     



       
\bibitem{jhep09}
      M.~Deak,   F.~Hautmann,     H.~Jung and K.~Kutak,     JHEP {\bf 0909} (2009) 121. 



\bibitem{pz_perugia}
      P.Z.~Skands,      arXiv:1005.3457 [hep-ph];   
      arXiv:0905.3418 [hep-ph],    in 
      Proceedings 1st     MPI Workshop (Perugia,  2008),  DESY-PROC-2009-06, 
      eds. P.~ Bartalini and   L.~Fan{\` o}.   

\bibitem{rdf1}
   R.D.~Field,   
   in Proceedings 1st
   MPI Workshop (Perugia,  2008),  DESY-PROC-2009-06, eds. P.~ Bartalini and 
   L.~Fan{\` o}.    

       
\bibitem{Sjostrand:2006za}
       T.~Sj{\" o}strand, S.~Mrenna, and P.~Skands, 
        JHEP {\bf 0605} (2006) 026. 
   

\bibitem{giese}
        M.~B{\" a}hr, S.~Gieseke and M.~Seymour,       JHEP {\bf 0807} (2008) 076.     


\bibitem{blok} 
          B.~Blok,   Yu.~Dokshitzer, L.~Frankfurt and M.~Strikman, 
          arXiv:1009.2714 [hep-ph]. 

\bibitem{strik-vogel} 
           M.~Strikman and W.~Vogelsang, arXiv:1009.6123 [hep-ph]. 
           
\bibitem{rog-strik} 
             T.C.~Rogers and    M.~Strikman,     Phys.\  Rev.\  D{\bf 81} (2010)   016013. 
                         
\bibitem{calu-trel} 
           G.~Calucci and D.~Treleani, arXiv:1009.5881 [hep-ph];   G.~Calucci, 
           talk at the 
          ISMD 2010            Symposium (Antwerp, September 2010).   
           
\bibitem{wiede_mpi}
          S.~Domdey, H.-J.~Pirner and U.A.~Wiedemann,  
           Eur.\ Phys.\ J.\  C {\bf 65} (2010)   153. 

\bibitem{maina_mpi} 
          E.~Maina,  
          arXiv:1010.5674 [hep-ph];   
             JHEP {\bf 0909} (2009) 081;     JHEP {\bf 0904} (2009) 098.  

\bibitem{berger_mpi} 
           E.L.~Berger,  C.B.~Jackson and G.~Shaughnessy,  
            Phys.\  Rev.\  D{\bf 81} (2010)   014014;  E.L.~Berger, talk 
            at ICHEP 2010 (Paris, July 2010).  
                         
\bibitem{gaunt} 
              J.R.~Gaunt and W.J.~Stirling,  JHEP {\bf 1003} (2010) 005; 
              J.R.~Gaunt, C.-H.~Kom,  
              A.~Kulesza and W.J.~Stirling,   
                 Eur.\ Phys.\ J.\  C {\bf 69} (2010)   53. 
     



\bibitem{proceed09}
      M.~Deak et al.,   
              arXiv:0908.1870 [hep-ph];   F.~Hautmann, arXiv:0909.1250 [hep-ph]. 


\bibitem{ch94}     
     S.~Catani and F.~Hautmann,   
     Nucl. Phys. B{\bf 427} (1994) 475;       Phys.  Lett. B{\bf 315}  (1993) 157. 
     

\bibitem{cj05}
           J.C.~Collins and H.~Jung, hep-ph/0508280. 
           
           
\bibitem{lundx}
               J.R.~Andersen et al,        Eur.\ Phys.\ J.\  C {\bf 48} (2006)   53.         

\bibitem{mueproc90c}
       A.H.~Mueller, Nucl.\ Phys.\  B  Proc.\ Suppl.\  {\bf 18C} 
       (1990) 125. 
  


\bibitem{forwjetcoeff}
      S.~Catani, M.~Ciafaloni and F.~Hautmann,    Nucl.\ Phys.\  B  Proc.\ Suppl.\  
       {\bf 29A}       (1992) 182. 


\bibitem{forwdis92}
      J.~Kwiecinski, A.D.~Martin and P.J.~Sutton,   Phys.\  Rev.\  D{\bf 46} (1992) 921; 
      J.~Bartels, A.~De Roeck and M.~Loewe, Z.\ Phys.\  {\bf  C54} (1992) 635; 
          W.K.~Tang,     Phys.\ Lett.\  B{\bf  278}  (1992) 363. 


\bibitem{laycock} 
        P.~Laycock et al.,  ``Future of DIS" summary report, in Proceedings of 
        the Workshop DIS 2010 (Florence, April 2010). 


\bibitem{cch-heraproc}
      S.~Catani, M.~Ciafaloni and F.~Hautmann,    in Proceedings of the Workshop 
      ``Physics at HERA",  Hamburg 1991, vol.~1, p.~690. 



\bibitem{ianmue} 
         E.~Iancu, M.S.~Kugeratski and D.N.~Triantafyllopoulos,   
             Nucl. Phys.  A{\bf 808}  (2008) 95;    
    E.~Iancu, C.~Marquet and G.~Soyez,
    Nucl.\ Phys.\  A {\bf 780}    (2006)  52;     C.~Marquet and R.~B.~Peschanski,
    Phys.\ Lett.\  B {\bf 587}   (2004)   201. 
   

\bibitem{hatta}   
              Y.~Hatta, E.~Iancu and A.H.~Mueller,          JHEP {\bf 0801} (2008) 026.  

\bibitem{gelis_etal_rvw} 
       F.~Gelis, T.~Lappi and R.~Venugopalan,    
       Int.\   J.\   Mod.\   Phys.\  E{\bf 16} (2007)  2595.
       

\bibitem{kov10} 
         W.A.~Horowitz and     Yu.V.~Kovchegov,     arXiv:1009.0545 [hep-ph]. 
             
\bibitem{weigert08}    
                 Yu.V.~Kovchegov and  H.~Weigert,       Nucl.\ Phys.\  A {\bf 807}    (2008)  158. 

\bibitem{gardi06} 
             E.~Gardi,  J.~Kuokkanen, K.~Rummukainen and H.~Weigert, 
                Nucl.\ Phys.\  A {\bf  784}    (2007)  282. 
                      

 \bibitem{kutak-absorpt}
                H.~Jung and K.~Kutak,    arXiv:0812.4082 [hep-ph]. 
         
\bibitem{avsar-stasto}
              E.~Avsar and A.M.~Stasto,     JHEP {\bf 1006} (2010) 112. 
   
 \bibitem{avsar-iancu}
              E.~Avsar and E.~Iancu,        Nucl.\ Phys.\  A {\bf 829}    (2009)  31. 
     

\bibitem{gelis_etal_hefnucl} 
       F.~Gelis, T.~Lappi and R.~Venugopalan,  
   Phys.\  Rev.\  D{\bf 78} (2008) 054019. 


\bibitem{hj_ang} 
       F.~Hautmann and H.~Jung,     
       JHEP {\bf 0810} (2008) 113. 
    
\bibitem{mw92}
     G.~Marchesini and B.R.~Webber,
     Nucl.\ Phys.\ {\bf B386} (1992) 215.

  
    
\bibitem{Catani:1989sg}
       S.~Catani, F.~Fiorani, and G.~Marchesini,    Nucl.\   Phys.\  {\bf B336} (1990)
           18.

\bibitem{skewang}
     M.~Ciafaloni, Nucl.\ Phys.\ {\bf B296} (1988)  49.  



   
\bibitem{hj_rec}     
        F.~Hautmann and H.~Jung,  Nucl.\ Phys.\  
   Proc.\ Suppl.\    {\bf 184}  (2008) 64 
   [arXiv:0712.0568 [hep-ph]]; 
   arXiv:0808.0873 [hep-ph].  

\bibitem{cascade_docu}
       H.~Jung et al.,      Eur.\  Phys.\   J.\  C {\bf 70} (2010) 1237  [arXiv:1008.0152 [hep-ph]]. 
       
\bibitem{cascade_comp}        
       H.~Jung,     Comput.\  Phys.\     Commun.\   {\bf 143} (2002) 100. 



 \bibitem{jadach09}
        S.~Jadach, A.~Kusina,     M.~Skrzypek and M.~Slawinska, 
          arXiv:1004.4131 [hep-ph];   arXiv:1002.0010 [hep-ph]; 
        S.~Jadach   and  M.~Skrzypek, arXiv:0909.5588 [hep-ph], 
         arXiv:0905.1399 [hep-ph].

\bibitem{watt_09}
       A.D.~Martin, M.G.~Ryskin and G.~Watt,     Eur.\ Phys.\ J.\  C {\bf 66} (2010)   163. 

\bibitem{gustafson} 
     C.~Flensburg and  G.~Gustafson,     arXiv:1004.5502 [hep-ph]; 
       G.~Gustafson,    Acta  Phys.\  Polon.\  B  {\bf 40} (2009) 1981. 



\bibitem{acta09}
     F.~Hautmann,   
     Acta  Phys.\  Polon.\  B  {\bf 40} (2009) 2139. 
     



   

\bibitem{hj_radcor}    
       F.~Hautmann and H.~Jung,            
         arXiv:0804.1746 [hep-ph], in Proc.\    8th International Symposium  on Radiative 
         Corrections   {\small RADCOR2007}  (Florence, October 2007). 


 

\bibitem{phot09}
        F.~Hautmann, arXiv:0909.1240 [hep-ph]. 

\bibitem{azim_bjets}        
         H.~Jung et al., 
        arXiv:1009.5067 [hep-ph]. 


\bibitem{deak_etal_higgs}
         M.~Deak et al., arXiv:1006.5401 [hep-ph];   
                    F.~Hautmann,  Phys.\ Lett.\ B {\bf 535} (2002) 159;  
                       F.~Hautmann,    H.~Jung and V.~Pandis,    arXiv:1011.6157 [hep-ph]. 

\bibitem{centr-jet-atl}
     ATLAS  Collaboration,    arXiv:1009.5908 [hep-ex].
           

\bibitem{centr-jet-cms}
         CMS Collaboration,    arXiv:1010.0203 [hep-ex].

  
\bibitem{jungetal_prep}
                  M.~Hentschinski,  private communication.   
 
\bibitem{Lai:1999wy}
                       H.L.~Lai et al.  
                      [CTEQ Collaboration],     Eur.\  Phys.\  J.\  {\bf C12} (2000)
                       375.  

\bibitem{fastjetpack} 
             M.~Cacciari and G.P.~Salam,      Phys.\ Lett.\ B {\bf 641} (2006) 57; 
             M.~Cacciari, G.P.~Salam and G.~Soyez,   http://fastjet.fr; 
             G.P.~Salam and G.~Soyez,        JHEP {\bf 0705}  (2007)   086. 


\bibitem{alioli} 
            S.~Alioli et al.,   arXiv:1012.3380 [hep-ph].    

\bibitem{bryan_eperp_10}
          A.~Papaefstathiou,   J.M.~Smillie and B.R.~Webber,        
               JHEP {\bf 1004}  (2010)   084. 

\bibitem{sung}
         I.~Sung,    Phys.\  Rev.\  D{\bf 80} (2009)   094020. 

\bibitem{yatta-ueda}
       Y.~Hatta and T.~Ueda,    Phys.\  Rev.\  D{\bf 80} (2009)   074018.


\bibitem{hef93}   
      S.~Catani, M.~Ciafaloni and F.~Hautmann,      
      Phys.\ Lett.\  B{\bf  307}  (1993) 147.    


\bibitem{becher-neub} 
          T.~Becher  and M.~Neubert,   arXiv:1007.4005  [hep-ph]; 
          S.~Mantry and F.~Petriello,  arXiv:1011.0757  [hep-ph]. 
               
\bibitem{rogers} 
        P.J.~Mulders and T.C.~Rogers,      
        Phys.\  Rev.\  D{\bf 81} (2010)   094006. 

\bibitem{idilbisci} 
          A.~Idilbi and I.~Scimemi, arXiv:1009.2776  [hep-ph]; 
           arXiv:1012.4419  [hep-ph]. 
        
\bibitem{chered} 
        I.~Cherednikov   and  N.~Stefanis,  
        Phys.\  Rev.\  D{\bf 80} (2009)   054008. 

\bibitem{bacch-diehl} 
           A.~Bacchetta,   D.~Boer,    M.~Diehl      and     P.J.~Mulders, 
               JHEP {\bf 0808} (2008) 023. 

\bibitem{rogers08} 
         T.C.~Rogers,      Phys.\  Rev.\  D{\bf 78} (2008)   074018. 


\bibitem{fhfeb07}
     F.~Hautmann,   Phys.\ Lett.\  B{\bf  655} (2007) 26; arXiv:0708.1319; 
    J.C.~Collins and F.~Hautmann,       JHEP {\bf 0103} (2001) 016; 
    Phys.\ Lett.\ B{\bf 472} (2000) 129. 

             
\bibitem{heraforw} 
        A.~Aktas et~al.,       Eur.\  Phys.\   J.\  C{\bf 46} (2006) 27; 
        S.~Chekanov et~al., Phys.  Lett.  B{\bf 632}  (2006) 13.


\bibitem{knut}  
        A.~Knutsson, LUNFD6-NFFL-7225-2007 (2007);  L.~J{\" o}nsson, 
         AIP Conf.\  Proc.\   828 (2006) 175. 

\bibitem{aaron-hera} 
        F.D.~Aaron et~al.,       Eur.\  Phys.\   J.\  C{\bf 54} (2008) 389. 


 
\end{thebibliography}\endgroup
\end{document}